\documentclass[12pt]{article}
\usepackage{geometry}
\usepackage{graphicx} 
\usepackage{caption}
\usepackage{subcaption}
\usepackage{stfloats}
\usepackage{amsmath,amsthm,amssymb,amsfonts,mathtools,bbold}
\usepackage[colorlinks=true
  ,urlcolor=blue
  ,anchorcolor=blue
  ,citecolor=blue
  ,filecolor=blue
  ,linkcolor=blue
  ,menucolor=blue
  ,linktocpage=true
  ,pdfproducer=medialab
  ,pdfa=true
]{hyperref}
\usepackage{xcolor}
\usepackage{enumerate}
\usepackage{cite}
\usepackage{color}
\usepackage{tikz}
\usetikzlibrary{decorations.pathreplacing}

\newcommand{\hs}{\mathcal{H}} 
\newcommand{\swap}{\mathcal{S}} 
\DeclareMathOperator{\tr}{tr} 
\newcommand{\B}{\mathcal{B}} 
\newcommand{\id}{\mathbb{1}} 

\newcommand\numberthis{\addtocounter{equation}{1}\tag{\theequation}}

\numberwithin{equation}{section}

\definecolor{Kblue}{HTML}{007Aff}
\definecolor{Korange}{HTML}{FF8800}
\definecolor{Kgreen}{HTML}{94E700}
\definecolor{Kred}{HTML}{FF0000}
\definecolor{Kpurple}{HTML}{82218b}

\begin{document}

\begin{titlepage}
{\ }
\vskip 1in

\begin{center}
{\LARGE Helping observers in closed universes \\[0.25cm] reach their full potential}
\vskip 0.5in 
{\large Kenneth Higginbotham\footnote{khigginbotham1@pitp.ca}}
\vskip 0.2in 
{\it Perimeter Institute for Theoretical Physics \\
Waterloo, Ontario N2L 2Y5, Canada}
\end{center}
\vskip 0.5in

\begin{abstract}\noindent
Recent work by Engelhardt, Gesteau, and Harlow applies proposals for incorporating observers into holographic maps to study the Antonini-Rath puzzle for closed universes. In a new form of ``observer complementarity,'' they find that an AdS bulk observer measures a SWAP test to determine that there is no closed universe in the bulk, contrary to the (limited) description given by an observer inside the closed universe. In this work, we improve the predictions of both observers by using the holographic maps to define new operators to perform this same SWAP test. With these, we show that the AdS observer cannot rule out a baby universe in the bulk, and the closed universe observer can improve the accuracy of their description.
\end{abstract}

\end{titlepage}

\tableofcontents

\section{Introduction} \label{sec:intro}

\begin{figure}
    \centering
    \begin{tikzpicture}

\newcommand{\AdS}[3]{
    \shade[bottom color=red, top color=red!60!black] 
        (#1,#2) ellipse [x radius=1.5cm, y radius=0.5cm];
    \draw[red!60!black, line width=3pt] 
        (#1,#2) ellipse [x radius=1.5cm, y radius=0.5cm];
    \node[scale=1.5,white] at (#1,#2) {$#3$};
    \node[scale=1.5,red!60!black] at (#1,#2) [shift={(1,-0.4)}] {$\uppercase{#3}$};
}

\newcommand{\babyPic}[2]{
    \def\ballcenter{(#1,#2)};
    \def\radius{1cm}; 
    \begin{scope}
        \clip \ballcenter circle [radius=\radius];
        \shade[inner color=blue!7!white, outer color=blue] 
            \ballcenter + (0.4,0.4) circle [radius=1.6cm]; 
    \end{scope}
    \draw[blue!60!black, thick] \ballcenter circle [radius=\radius];
    \draw[thick, blue]
        \ballcenter + (\radius,0)
        [xscale=1, yscale=0.4]
        arc[start angle=0, end angle=-180, radius=1];
    \node[scale=1.5,white] at \ballcenter [shift={(-0.3,0.05)}] {$b$};
}

\node[scale=1.4] at (-8,3) {$|\psi_1\rangle$};
\AdS{-4}{3}{a};
\babyPic{0}{3};
\AdS{4}{3}{a};

\node[scale=1.4] at (-8,0) {$|\psi_2\rangle$};
\AdS{-4}{0}{a};
\AdS{4}{0}{a};

\end{tikzpicture}
    \caption{The two candidate bulk states described by AR \cite{antonini_holographic_2025}. The top line, denoted $|\psi_1\rangle$, depicts ``description 1'' prepared by the AS$^2$ construction \cite{antonini_cosmology_2023}, consisting of a baby universe $b$ and two disconnected AdS spacetimes $a$. The bottom line, denoted $|\psi_2\rangle$, depicts ``description 2'' without the baby universe. Figure adapted from \cite{higginbotham_tests_2025}.}
    \label{fig:AS2_AR}
\end{figure}
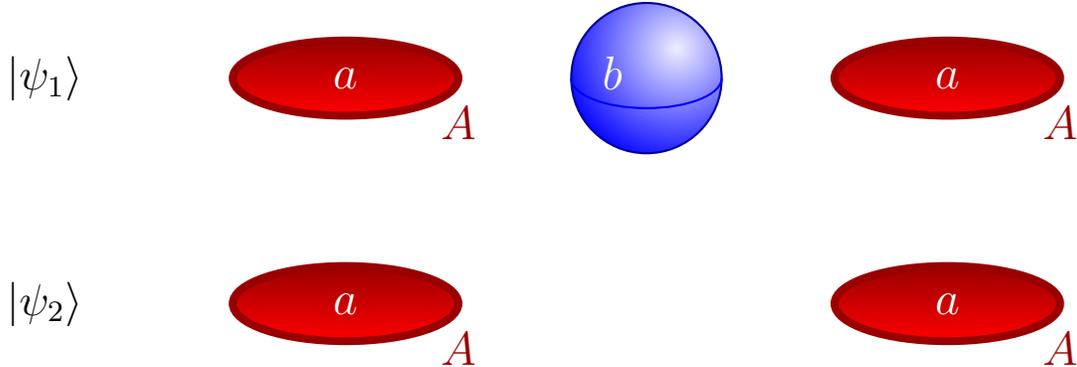

Understanding the nature of closed universes in quantum gravity has long been an exciting and challenging problem.\footnote{The term ``baby universe'' is also commonly used, particularly in contexts where closed universes appear alongside a ``parent universe'' such as in the gravitational path integral. We will use both terms interchangeably here.} Recent progress has been motivated by two puzzling observations:
\begin{enumerate}
    \item \textit{Trivial Hilbert space:} evidence suggests that the fundamental Hilbert space of a closed universe is 1-dimensional \cite{maldacena_Wormholes_2004,
    almheiri_Page_2020,
    penington_Replica_2020,
    marolf_Transcending_2020,
    mcnamara_Baby_2020,
    usatyuk_Closed_2024,
    usatyuk_Closed_2025}.
    
    \item \textit{Antonini-Rath (AR) puzzle:} two bulk states -- one with a closed universe \cite{antonini_cosmology_2023} and one without, depicted in figure \ref{fig:AS2_AR} -- appear to have the same boundary CFT dual \cite{antonini_holographic_2025}.
\end{enumerate}
These puzzles are sharpest in the Antonini-Sasieta-Swingle (AS$^2$) \cite{antonini_cosmology_2023} construction for a closed universe in AdS/CFT; see \cite{antonini_baby_2025,sasieta_baby_2025,belin_baby_2025} for recent studies of this and related constructions. Attempts to resolve both puzzles fall broadly into two categories originally proposed by AR in \cite{antonini_holographic_2025}. The first calls into question the validity of semiclassical closed universes; such work includes a bulk SWAP test proposed by Engelhardt and Gesteau (EG) \cite{engelhardt_further_2025} and a no-go theorem by Gesteau \cite{gesteau_nogo_2025}. The second suggests that something beyond traditional AdS/CFT \cite{maldacena_Large_1999,gubser_Gauge_1998,witten_Sitter_1998} is required to describe closed universes holographically; these include incorporating observers into gravitational path integrals and non-isometric codes \cite{harlow_quantum_2025,abdalla_Gravitational_2025,akers_observers_2025,chen_Observers_2025} as well as averaged large $N$ limits \cite{liu_holographic_2025,kudler-flam_emergent_2025,liu_filtering_2025}.

A recent proposal by Engelhardt, Gesteau, and Harlow (EGH) \cite{engelhardt_observer_2025} combines both EG's SWAP test and the above observer work into a new form of ``observer complementarity'' for closed universes. EGH considers two bulk observers in the AS$^2$ construction \cite{antonini_cosmology_2023}: one in the bulk AdS spacetime (denoted $\alpha$) and one in the baby universe (denoted $\beta$). By constructing new holographic maps for $\alpha$ and $\beta$ according to the observer rules proposed by Harlow-Usatyuk-Zhao (HUZ) \cite{harlow_quantum_2025}, EGH makes predictions for how each observer would measure the bulk SWAP test. They find that $\alpha$'s measurement is consistent with \textit{no} baby universe present in the bulk, while $\beta$'s description must include the baby universe in which they live. In the spirit of complementarity \cite{susskind_stretched_1993}, EGH argues that the two observers are allowed to disagree on the existence of the baby universe because they are never in causal contact.

Even though $\beta$ can describe the baby universe, EGH finds that the accuracy of their description is limited by their own entropy. These results are consistent with EG's prior claim in \cite{engelhardt_further_2025} that the SWAP test does not support a valid semiclassical description of the closed universe -- only $\beta$ sees the closed universe, and even their description is limited. However, the ability of the SWAP test to rule out a semiclassical closed universe has since been challenged \cite{higginbotham_tests_2025,antonini_baby_2025}. 

In \cite{higginbotham_tests_2025}, we found that the holographic maps naturally define a new operator to perform the SWAP test when the bulk contains a baby universe. We demonstrated that a measurement of this operator, which includes a projector acting on the baby universe, exactly matches a measurement of the SWAP test without the baby universe. Therefore, the SWAP test cannot be used to distinguish between the two candidate bulk states, restoring some confidence in the semiclassical description of baby universes. 

In this work, we aim to take the same care with the observers considered by EGH. We do this in two ways. First, we use holographic maps constructed from both the HUZ \cite{harlow_quantum_2025} and Colorado (CO) \cite{akers_observers_2025} observer rules to define new operators performing the SWAP test to greater accuracy in each observer's description. Second, we find that averaging can pick up contributions from holographic maps for which the AR puzzle does not hold. When it is important to preserve the puzzle, we consider toy models that avoid averaging.

With both of these improvements, we find that the $\alpha$ observer cannot use the SWAP test to rule out a bulk description with a baby universe; instead, they can equally well describe the bulk with or without one. While $\beta$'s prediction for the SWAP test is still generically limited, we find that the accuracy of their prediction can be improved. This improvement is controlled by entanglement between the baby universe and AdS spacetimes, consistent with results from \cite{antonini_cosmology_2023,antonini_baby_2025}. Furthermore, we find that $\beta$'s new operator reproduces the SWAP test \textit{perfectly} in a toy model, demonstrating that this limitation in $\beta$'s accuracy is not fundamental.

The remainder of this work will be structured as follows. Section \ref{sec:web} will review the holographic maps defined by EGH for bulk Hilbert spaces with and without the baby universe and observers. Next, section \ref{sec:swap} will define new operators performing the SWAP test on each Hilbert space using tools inspired by \cite{akers_reconstruction_2025}. Section \ref{sec:HUZ} will specialize to the HUZ rules for $\alpha$ and $\beta$; the results of this section should be directly comparable with EGH's work in \cite{engelhardt_observer_2025}. Section \ref{sec:CO} goes further to consider the CO rules for $\alpha$ and $\beta$. Finally, we summarize our results (see tables \ref{tab:summary} and \ref{tab:beta}) and discuss open questions in section \ref{sec:conc}.

\section{A web of holographic maps} \label{sec:web}

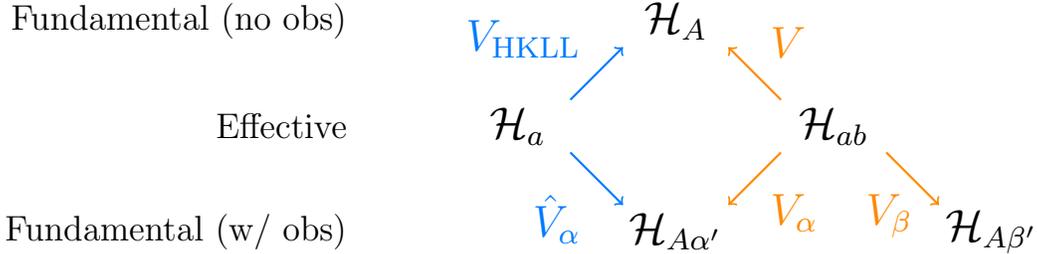
\begin{figure}[t]
    \centering
    \begin{tikzpicture}[thick,scale=1.4, every node/.style={transform shape}]
    
\node[scale=0.8, anchor=east] at (1.5,1) {Fundamental (no obs)};
\node[scale=0.8, anchor=east] at (1.5,0) {Effective};
\node[scale=0.8, anchor=east] at (1.5,-1) {Fundamental (w/ obs)};

\node at (3,0) {$\hs_a$};
\node at (4.5,1) {$\hs_A$};
\node at (4.5,-1) {$\hs_{A\alpha'}$};
\node at (6,0) {$\hs_{ab}$};
\node at (7.5,-1) {$\hs_{A\beta'}$};

\draw[Kblue] (3.5,0.25) edge[->] (4,0.75);
\draw[Kblue] (3.5,-0.25) edge[->] (4,-0.75);
\draw[Korange] (5.5,0.25) edge[->] (5,0.75);
\draw[Korange] (5.5,-0.25) edge[->] (5,-0.75);
\draw[Korange] (6.5,-0.25) edge[->] (7,-0.75);

\node[Kblue, anchor=south east] at (3.75,0.5) {$V_\text{HKLL}$};
\node[Kblue, anchor=north east] at (3.75,-0.5) {$\hat{V}_\alpha$};
\node[Korange, anchor=south west] at (5.25,0.5) {$V$};
\node[Korange, anchor=north west] at (5.25,-0.5) {$V_\alpha$};
\node[Korange, anchor=north east] at (6.9,-0.5) {$V_\beta$};
    
\end{tikzpicture}
    \caption{A diagram representing the five holographic maps defined in \cite{engelhardt_observer_2025} and how they relate Hilbert spaces with and without the baby universe ($b$) and observers ($\alpha$, $\beta$). The middle row contains effective Hilbert spaces, the top row contains fundamental Hilbert spaces without any observer, and the bottom row contains fundamental Hilbert spaces with one observer. Holographic maps are color coded: \textcolor{Kblue}{blue} for isometries, and \textcolor{Korange}{orange} for non-isometries.}
    \label{fig:HS_map}
\end{figure}

We begin with a review of the five holographic maps defined by EGH in \cite{engelhardt_observer_2025} between two bulk (\textit{effective}) and three boundary (\textit{fundamental}) Hilbert spaces. Figure \ref{fig:HS_map} will serve as a helpful guide throughout this section. Notation for the AS$^2$ construction \cite{antonini_cosmology_2023} and AR's candidate bulk states \cite{antonini_holographic_2025} on which these maps are based differs slightly across \cite{antonini_cosmology_2023,antonini_holographic_2025,engelhardt_further_2025,engelhardt_observer_2025,higginbotham_tests_2025,antonini_baby_2025}. For ease of comparison with EGH's work in \cite{engelhardt_observer_2025}, we will adopt their notation here:
\begin{align*}
    a &\equiv \text{two disconnected AdS spacetimes} \\[0.2cm]
    b &\equiv \text{baby universe} \\[0.2cm]
    A &\equiv \text{two disconnected AdS boundaries, } A=\partial a \\[0.2cm]
    \hs_a &\equiv \text{bulk (\textit{effective}) Hilbert space on AdS timeslice} \\[0.2cm]
    \hs_b &\equiv \text{bulk (\textit{effective}) Hilbert space on baby universe timeslice} \\[0.2cm]
    \hs_A &\equiv \text{boundary (\textit{fundamental}) Hilbert space} \\[0.2cm]
    |\psi_1\rangle &\equiv \text{AR ``description 1'' with baby universe, state in } \hs_{ab} = \hs_a \otimes \hs_b \\[0.2cm]
    |\psi_2\rangle &\equiv \text{AR ``description 2'' without baby universe, state in } \hs_a
\end{align*}
Figure \ref{fig:AS2_AR} provides an illustration of the two bulk states $|\psi_1\rangle$ and $|\psi_2\rangle$.

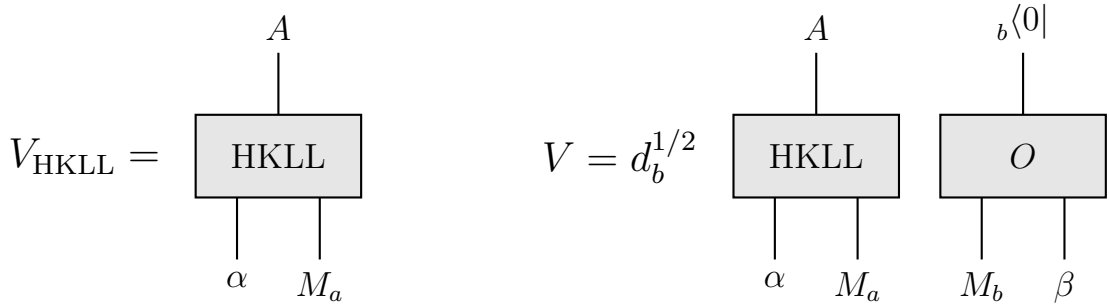
\begin{figure}
    \centering
    \begin{tikzpicture}[thick,scale=1.1, every node/.style={transform shape}]

\draw (1,0.25) -- (1,1);
\node[anchor=north] at (1,0.25) {$M_a$};
\draw (0,0.25) -- (0,1);
\node[anchor=north] at (0,0.25) {$\alpha$};
\draw[fill=gray!20] (-0.5,1) rectangle (1.5,2);
\node at (0.5,1.5) {HKLL};
\draw (0.5,2) -- (0.5,2.75);
\node[anchor=south] at (0.5,2.75) {$A$};
\node[anchor=east,scale=1.2] at (-0.75, 1.5) {$V_\text{HKLL} =$};

\begin{scope}[shift={(9,0)}]
    \draw (1,0.25) -- (1,1);
    \node[anchor=north] at (1,0.25) {$\beta$};
    \draw (-1.5,0.25) -- (-1.5,1);
    \node[anchor=north] at (-1.5,0.25) {$M_a$};
    \draw[fill=gray!20] (-3,1) rectangle (-1,2);
    \node at (-2,1.5) {HKLL};
    \draw (0.5,2) -- (0.5,2.75);
    \node[anchor=south] at (0.5,2.75) {$\prescript{}{b}{\langle0|}$};
    \draw (0,0.25) -- (0,1);
    \node[anchor=north] at (0,0.25) {$M_b$};
    \draw (-2.5,0.25) -- (-2.5,1);
    \node[anchor=north] at (-2.5,0.25) {$\alpha$};
    \draw[fill=gray!20] (-0.5,1) rectangle (1.5,2);
    \node at (0.5,1.5) {$O$};
    \draw (-2,2) -- (-2,2.75);
    \node[anchor=south] at (-2,2.75) {$A$};
    \node[anchor=east,scale=1.2] at (-3.25, 1.5) {$V = d_b^{1/2}$};
\end{scope}
        
\end{tikzpicture}
    \caption{Circuit diagrams for holographic maps without an observer. $V_\text{HKLL}$ (left) maps a bulk state without a baby universe to the asymptotic AdS boundary; defined from the HKLL map, it is isometric. $V$ (right) includes post-selection on the baby universe and is non-isometric. Bulk inputs have been split into subsystems for observers ($\alpha$, $\beta$) and matter ($M_a$, $M_b$) for later use in applying observer rules.}
    \label{fig:map2A}
\end{figure}

\paragraph{No observer.} When no observer is present in the bulk, holographic maps are defined in the usual way. Without a baby universe, EGH uses the HKLL map \cite{hamilton_local_2006,hamilton_holographic_2006,hamilton_local_2007} to define a holographic map acting on the AdS spacetimes,
\begin{equation}
    V_\text{HKLL}: \hs_a \to \hs_A,
\end{equation}
which is isometric. Because the baby universe has no asymptotic boundary, it must be mapped to the same fundamental Hilbert space. EGH denotes such a holographic map for the bulk with a baby universe as $V$,
\begin{equation}
    V: \hs_{ab} \to \hs_A.
\end{equation}
Since the $\hs_a$ subsystem is common to both effective Hilbert spaces, $V$ must act as $V_\text{HKLL}$ on $\hs_a \subset \hs_{ab}$ while post-selecting on the entirety of $\hs_b \subset \hs_{ab}$ \cite{engelhardt_observer_2025,higginbotham_tests_2025}, 
\begin{equation} \label{eq:V}
    V = d_b^{1/2} \, V_\text{HKLL} \otimes \prescript{}{b}{\langle}0| O.
\end{equation}
$V$ is non-isometric due to the post-selection on the baby universe. The prefactor $d_b \equiv \dim \hs_b$ is included to preserve state normalization for average choices of the orthogonal $O \in \B(\hs_b)$.\footnote{An orthogonal operator was chosen instead of a unitary operator in the definition of $V$ because gauged $\mathcal{CRT}$ invariance implies the Hilbert space of a closed universe is real \cite{harlow_gauging_2023}.} Both $V_\text{HKLL}$ and $V$ are included in the top half of figure \ref{fig:HS_map}, and circuit diagrams for both are shown in figure \ref{fig:map2A}.

\paragraph{One observer.} In order to incorporate an observer, we must first identify the observer as a subsystem of the bulk effective Hilbert space. The $\alpha$ observer in the AdS bulk is identified as a subsystem of $\hs_a$,
\begin{equation} \label{eq:alpha_sub}
    \hs_a = \hs_\alpha \otimes \hs_{M_a}
\end{equation}
where $M_a$ denotes matter in the remainder of the AdS bulk that constitutes the $\alpha$ observer's environment. Similarly, the $\beta$ observer is identified as a subsystem of $\hs_b$,
\begin{equation} \label{eq:beta_sub}
    \hs_b = \hs_\beta \otimes \hs_{M_b}
\end{equation}
with matter in the baby universe denoted by $M_b$. These subsystems have been identified in the inputs to the circuit diagrams for $V_\text{HKLL}$ and $V$ in figure \ref{fig:map2A}.

Once an observer subsystem has been identified, we modify the holographic map to treat the chosen observer in a special way. Either the HUZ \cite{harlow_quantum_2025} or CO \cite{akers_observers_2025} rules can be used to make these modifications. We do not make a particular choice here; sections \ref{sec:HUZ} and \ref{sec:CO} will specialize to the HUZ and CO rules, respectively. Until then, we will refer generically to the ``observer rules'' or ``observer-modified holographic maps''. The reader may refer to figures \ref{fig:HUZ_rules} and \ref{fig:CO_rules} if they wish for a visual example during the remainder of this section.

Generically, the observer-modified holographic maps will result in new fundamental Hilbert spaces that are larger by an observer-dependent factor. For example, consider the $\alpha$ observer in a bulk with no baby universe. In this case, the observer rules for $\alpha$ must be applied to $V_\text{HKLL}$. EGH denotes this new holographic map as $\hat{V}_\alpha$, and we denote the enlarged fundamental Hilbert space as $\hs_{A \alpha'}$:
\begin{equation}
    \hat{V}_\alpha: \hs_a \to \hs_{A\alpha'}.
\end{equation}
The $\alpha$ observer living in an AdS bulk accompanied by a baby universe is treated in a similar way. This time, the observer modification is applied to $V$; EGH denotes this new holographic map as $V_\alpha$. Again, since the baby universe has no asymptotic boundary, $V_\alpha$ maps $\hs_{ab}$ to the same fundamental Hilbert space as $\hat{V}_\alpha$,
\begin{equation}
    V_\alpha: \hs_{ab} \to \hs_{A\alpha'}.
\end{equation}
Just as in (\ref{eq:V}), $V_\alpha$ and $\hat{V}_\alpha$ must be related as
\begin{equation} \label{eq:V_alpha}
    V_\alpha = d_b^{1/2} \, \hat{V}_\alpha \otimes \prescript{}{b}{\langle 0|} O
\end{equation}
since both holographic maps must act in the same way on their common $\hs_a$ subsystem.

Finally, taking $\beta$ in the closed universe to be an observer amounts to applying the observer rules to the $\beta$ subsystem in $V$. EGH denotes this final holographic map as $V_\beta$, and we denote the resulting fundamental Hilbert space as $\hs_{A\beta'}$,
\begin{equation}
    V_\beta: \hs_{ab} \to \hs_{A\beta'}.
\end{equation}
All three of these observer-modified maps ($\hat{V}_\alpha$, $V_\alpha$, and $V_\beta$) are shown in the bottom half of figure \ref{fig:HS_map}, completing the web of holographic maps. Note that of these three only $\hat{V}_\alpha$ is isometric since it alone was derived from the isometric $V_\text{HKLL}$ map. Specific constructions for $\hat{V}_\alpha$, $V_\alpha$, and $V_\beta$ will be given using the HUZ rules in section \ref{sec:HUZ} and CO rules in section \ref{sec:CO}.

\paragraph{Bulk states.} Given the subdivisions of $\hs_a$ and $\hs_b$ into observer and matter subsystems in (\ref{eq:alpha_sub}) and (\ref{eq:beta_sub}), let us refine our definitions of $|\psi_1\rangle$ and $|\psi_2\rangle$. Throughout, we will follow EGH in taking the $\alpha$ and $\beta$ subsystems to be in a pure state; this will simplify the definition of the SWAP test in the next section. We therefore define ``description 1'' with the baby universe to be given by the state
\begin{equation} \label{eq:psi1}
    |\psi_1\rangle \equiv |\psi_1\rangle_{M_aM_b} |\psi_\alpha\rangle |\psi_\beta\rangle \in \hs_{ab}.
\end{equation}
Following the construction of the AS$^2$ geometry, $|\psi_1\rangle_{M_aM_b}$ is generically entangled between $M_a$ and $M_b$. Similarly, ``description 2'' will be given by the state
\begin{equation} \label{eq:psi2}
    |\psi_2\rangle \equiv |\psi_2\rangle_{M_a} |\psi_\alpha\rangle \in \hs_a.
\end{equation}
Because there is no baby universe in description 2, the AdS matter $M_a$ must be in a pure state, here given by $|\psi_2\rangle_{M_a}$.

Equipped with these holographic maps and bulk states, we express the AR puzzle as \cite{engelhardt_observer_2025,higginbotham_tests_2025}:
\begin{equation} \label{eq:ARcond}
    V |\psi_1\rangle = V_\text{HKLL} |\psi_2\rangle.
\end{equation}
Throughout, we will refer to (\ref{eq:ARcond}) as the ``AR condition''. Note that given a particular choice of $V$, it is only possible to satisfy the AR condition if $|\psi_1\rangle$ is not in the kernel of $V$. Alternatively, given particular bulk states $|\psi_1\rangle$ and $|\psi_2\rangle$, the post-selection in $V$ can be chosen to satisfy (\ref{eq:ARcond}).

\section{The SWAP test} \label{sec:swap}

We now use the holographic maps reviewed in section \ref{sec:web} to define operators performing the SWAP test on each of the five Hilbert spaces in figure \ref{fig:HS_map}. We begin by reviewing some general properties of linear maps that will be useful for these definitions, following \cite{akers_reconstruction_2025}.

\subsection{Linear maps for states and operators} \label{sec:maps}
Let us consider a generic linear map $V$ from some Hilbert space $\hs_1$ to a second Hilbert space $\hs_2$,
\begin{equation}
    V: \mathcal{H}_1 \to \mathcal{H}_2.
\end{equation}
We define $V$ to map a state of the first Hilbert space $\rho_1 \in \hs_1$ to $\rho_2 \in \hs_2$ as
\begin{equation} \label{eq:Vrho1}
    V: \rho_1 \to \rho_2 = V \rho_1 V^\dagger.
\end{equation}
Now, consider the expectation value of some operator $O_2 \in \B(\hs_2)$ in state $\rho_2$. Using $V$, we may rewrite this expectation value in terms of $\rho_1$ as
\begin{equation} \label{eq:V*exp}
    \langle O_2 \rangle_{\rho_2} = \text{tr} \big( \rho_2 O_2 \big) = \text{tr} \big( V \rho_1 V^\dagger O_2 \big) = \text{tr} \big( \rho_1 V^\dagger O_2 V \big) = \text{tr} \big( \rho_1 O_1 \big) = \langle O_1 \rangle_{\rho_1}
\end{equation}
where we have used (\ref{eq:Vrho1}) in the second equality and the cyclicity of the trace in the third equality. In the fourth equality, we have defined
\begin{equation} \label{eq:O1}
    O_1 \equiv V^\dagger O_2 V \in \mathcal{B}(\mathcal{H}_1).
\end{equation}
The map $V$ on Hilbert spaces has induced a map on operators that we will denote as $V^*: \B(\hs_2) \to \B(\hs_1)$ following \cite{akers_reconstruction_2025}. We take (\ref{eq:O1}) to define the action of $V^*$ on operators in $\B(\hs_2)$,
\begin{equation}
    V^*: O_2 \to O_1 = V^\dagger O_2 V.
\end{equation}
We can think of $V^*$ as the ``Heisenberg picture'' associated with the ``Schr\"odinger picture'' given by $V$.

There are two important facts to note about this induced operator map $V^*$. First, (\ref{eq:V*exp}) demonstrates that $V^*$ preserves the expectation value of $O_2$ on states related by $V$, $\langle O_2 \rangle_{\rho_2} = \langle O_1 \rangle_{\rho_1}$. Second, nowhere in (\ref{eq:V*exp}) did we require $V$ to be isometric such that $V^\dagger V = \id$. The existence of $V^*$ holds for both isometric and non-isometric maps; $V$ \textit{always} induces $V^*$.

Suppose instead that we are first given the operator $O_1\in\B(\hs_1)$ and asked about the corresponding operator acting on $\hs_2$. In this case $V^*$ is not very helpful. Instead, we need a map that takes operators in $\B(\hs_1)$ to corresponding operators in $\B(\hs_2)$. Following \cite{akers_reconstruction_2025}, we denote this map as $R^*$,
\begin{equation} \label{eq:R*}
    R^*: \mathcal{B}(\mathcal{H}_1) \to \mathcal{B}(\mathcal{H}_2),
\end{equation}
and call $R^*(O_1)\in\B(\hs_2)$ the ``reconstruction'' of $O_1$ on $\hs_2$. While $R^*$ should be inverted by $V^*$, 
\begin{equation} \label{eq:invert}
    V^* R^* : O_1 \to O_1,
\end{equation}
it is not induced by $V$ like $V^*$ is. However, a natural definition of this reconstruction map $R^*$ does exist when $V$ is isometric,
\begin{equation} \label{eq:natural_R*}
    R^*: O_1 \to O_2 = V O_1 V^\dagger, \qquad V \text{ isometric.}
\end{equation}
The isometry of $V$ guarantees that this definition of $R^*$ is inverted by $V^*$ defined in (\ref{eq:O1}). 

If $V$ is non-isometric, there is no such natural definition of $R^*$. In fact, there likely does not exist a reconstruction map $R^*$ that satisfies (\ref{eq:invert}) exactly for a given choice of $O_1$. In other words, there might not exist an operator in $\B(\hs_2)$ that maps exactly to $O_1$ under $V^*$. The best we can do in this case is to construct an $R^*$ such that $V^*R^*$ maps $O_1$ to another operator in $\B(\hs_1)$ that is close in some measure. As a result, the expectation value of $O_1$ on $\rho_1$ will only approximately be preserved by its reconstruction on $\hs_2$,
\begin{equation}
    \langle O_1 \rangle_{\rho_1} \approx \langle R^*(O_1)\rangle_{V \rho_1 V^\dagger}, \qquad V \text{ non-isometric.}
\end{equation}

In the following, $V$ will be given by one of the five holographic maps from a bulk (\textit{effective}) Hilbert space to a boundary (\textit{fundamental}) Hilbert space. The induced operator map $V^*$ then maps boundary operators to bulk operators, while the (possibly approximate) reconstruction map $R^*$ maps bulk operators to boundary operators.

\subsection{Defining new operators} \label{sec:newops}
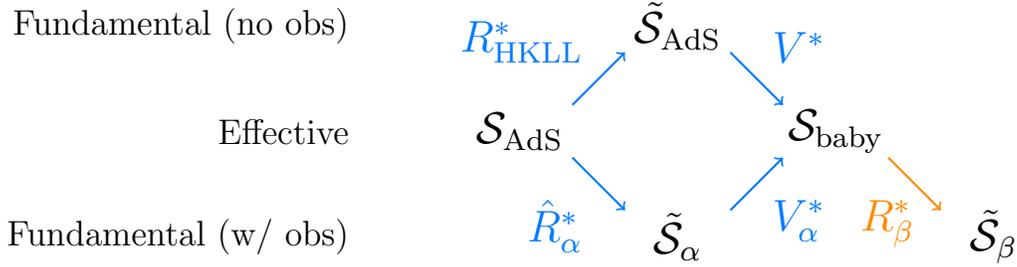
\begin{figure}
    \centering
    \begin{tikzpicture}[thick,scale=1.4, every node/.style={transform shape}]

\node[scale=0.8, anchor=east] at (1.5,1) {Fundamental (no obs)};
\node[scale=0.8, anchor=east] at (1.5,0) {Effective};
\node[scale=0.8, anchor=east] at (1.5,-1) {Fundamental (w/ obs)};

\node at (3,0) {$\swap_\text{AdS}$};
\node at (4.5,1) {$\tilde{\swap}_\text{AdS}$};
\node at (4.5,-1) {$\tilde{\swap}_{\alpha}$};
\node at (6,0) {$\swap_\text{baby}$};
\node at (7.5,-1) {$\tilde{\swap}_{\beta}$};

\draw[Kblue] (3.5,0.25) edge[->] (4,0.75);
\draw[Kblue] (3.5,-0.25) edge[->] (4,-0.75);
\draw[Kblue] (5,0.75) edge[->] (5.5,0.25);
\draw[Kblue] (5,-0.75) edge[->] (5.5,-0.25);
\draw[Korange] (6.5,-0.25) edge[->] (7,-0.75);

\node[Kblue, anchor=south east] at (3.75,0.5) {$R^*_\text{HKLL}$};
\node[Kblue, anchor=north east] at (3.75,-0.5) {$\hat{R}^*_\alpha$};
\node[Kblue, anchor=south west] at (5.25,0.5) {$V^*$};
\node[Kblue, anchor=north west] at (5.25,-0.5) {$V^*_\alpha$};
\node[Korange, anchor=north east] at (6.9,-0.5) {$R^*_\beta$};

\end{tikzpicture}
    \caption{A diagram representing the operators performing the SWAP test on the Hilbert spaces of figure \ref{fig:HS_map}. Operators on fundamental Hilbert spaces have been distinguished with a tilde. Arrows indicate how each is defined relative to $\swap_\text{AdS}$ and are colored according to whether the resulting operator performs the SWAP test exactly (\textcolor{Kblue}{blue}) or approximately (\textcolor{Korange}{orange}).}
    \label{fig:Op_map}
\end{figure}

Let us now apply these tools to define operators implementing the SWAP test on the web of Hilbert spaces in figure \ref{fig:HS_map}. Figure \ref{fig:Op_map} will serve as a helpful guide throughout the remainder of this section.

The SWAP test was originally defined in \cite{engelhardt_further_2025} to swap the bulk state on two copies of $\hs_a$. With the introduction of the $\alpha$ observer in the pure state $|\psi_\alpha\rangle$, EGH defines the operator $\swap_\text{AdS}$ acting on two copies of $\hs_a$ to only swap the bulk matter $M_a$ while acting as the identity on $\alpha$:
\begin{equation} \label{eq:S_AdS}
     \swap_\text{AdS} \equiv \text{SWAP}_{M_a} \otimes \id_\alpha.
\end{equation}
Since the matter $M_a$ in ``description 2'' given by (\ref{eq:psi2}) is taken to be in a pure state, we immediately find that the expectation value of $\swap_\text{AdS}$ on $|\psi_2\rangle$ is
\begin{equation} \label{eq:expSads}
    \langle \swap_\text{AdS} \rangle_{|\psi_2\rangle} = 1.
\end{equation}
We use this operator as the starting point for defining the SWAP test on the remaining four Hilbert spaces.

Fortunately, the two holographic maps $V_\text{HKLL}$ and $\hat{V}_\alpha$ acting on $\hs_a$ are isometric. This allows us to naturally define reconstruction maps using (\ref{eq:natural_R*}) that can map $\swap_\text{AdS}$ to the fundamental Hilbert spaces $\hs_A$ and $\hs_{A\alpha'}$. EGH provides exactly this definition of the SWAP test acting on two copies of $\hs_A$ via the HKLL map,\footnote{To be precise, $\swap_\text{AdS}$ should transform under two copies of $V_\text{HKLL}$. To keep our expressions readable, we will abuse notation and only write one copy of $V_\text{HKLL}$.}
\begin{equation} \label{eq:SAdS_recon}
    \tilde{\swap}_\text{AdS} \equiv V_\text{HKLL} \swap_\text{AdS} V^\dagger_\text{HKLL}
\end{equation}
where a tilde has been used to distinguish this as an operator acting on a fundamental Hilbert space. Following (\ref{eq:natural_R*}), we take this as the definition of a reconstruction map $R^*_\text{HKLL}$,
\begin{equation} 
    R^*_\text{HKLL}: \swap_\text{AdS} \to \tilde{\swap}_\text{AdS}.
\end{equation}
In the exact same way, the isometry of $\hat{V}_\alpha$ permits a natural reconstruction of $\swap_\text{AdS}$ on two copies of $\hs_{A\alpha'}$. We denote this reconstruction map as $\hat{R}^*_\alpha$ and its reconstruction of $\swap_\text{AdS}$ as $\tilde\swap_\alpha$: 
\begin{equation} \label{eq:S_alpha}
    \hat{R}^*_\alpha : \swap_\text{AdS} \to \tilde{\swap}_\alpha \equiv \hat{V}_\alpha \swap_\text{AdS} \hat{V}^\dagger_\alpha.
\end{equation}
The isometry of $V_\text{HKLL}$ and $\hat{V}_\alpha$ further implies that both reconstructions $\tilde\swap_\text{AdS}$ and $\tilde\swap_\alpha$ have the same expectation value as $\swap_\text{AdS}$ on states related to $|\psi_2\rangle$:
\begin{equation} \label{eq:exp_psi2}
    \langle \tilde{\swap}_\alpha \rangle_{\hat{V}_\alpha |\psi_2\rangle} = \langle \tilde{\swap}_\text{AdS} \rangle_{V_\text{HKLL} |\psi_2\rangle} = \langle \swap_\text{AdS} \rangle_{|\psi_2\rangle} = 1.
\end{equation}
We therefore interpret $\tilde\swap_\alpha$ as the correct way for the $\alpha$ observer to perform the SWAP test. These three SWAP test operators ($\swap_\text{AdS}$, $\tilde\swap_\text{AdS}$, and $\tilde\swap_\alpha$) as well as their reconstruction maps ($R^*_\text{HKLL}$ and $\hat{R}^*_\alpha$) are shown on the left side of figure \ref{fig:Op_map}. 

Next, consider how the SWAP test should act on two copies of the bulk Hilbert space $\hs_{ab}$ with a baby universe. Recall that even though both $V$ and $V_\alpha$ acting on $\hs_{ab}$ are non-isometric, they still induce operator maps $V^*$ and $V^*_\alpha$. Therefore, we can define the SWAP test for the bulk with a baby universe by applying the induced operator map $V^*$ to the fundamental operator $\tilde{\swap}_\text{AdS}$ already defined. We denote this new operator as $\swap_\text{baby}$:\footnote{This is the same operator we previously defined in \cite{higginbotham_tests_2025}. There, we interpreted $\swap_\text{baby}$ as the ``non-perturbatively correct'' version of the ``perturbative'' operator $\swap_\text{AdS}\otimes\id_b$ used to perform the SWAP test in \cite{engelhardt_further_2025}.}
\begin{align*} 
    V^*: \tilde{\swap}_\text{AdS} \to \swap_\text{baby} &\equiv V^\dagger \tilde{\swap}_\text{AdS} V \\
        &= V^\dagger V_\text{HKLL} \swap_\text{AdS} V^\dagger_\text{HKLL} V \\
        &= d_b \, \swap_\text{AdS} \otimes O^T |0\rangle\langle0| O, \numberthis \label{eq:S_baby}
\end{align*}
where we used (\ref{eq:SAdS_recon}) in the second line and (\ref{eq:V}) with the isometry of $V_\text{HKLL}$ in the last line. Instead, we could have used $V^*_\alpha$ to define $\swap_\text{baby}$ from $\tilde\swap_\alpha$ as
\begin{align*} 
    V^*_\alpha: \tilde{\swap}_\alpha \to \swap_\text{baby} &\equiv V^\dagger_\alpha \tilde{\swap}_\alpha V_\alpha \\
        &= V^\dagger_\alpha \hat{V}_\alpha \swap_\text{AdS} \hat{V}^\dagger_\alpha V_\alpha \\
        &= d_b \, \swap_\text{AdS} \otimes O^T |0\rangle\langle0| O, \numberthis \label{eq:S_baby2}
\end{align*}
where we used (\ref{eq:S_alpha}) in the second line and (\ref{eq:V_alpha}) with the isometry of $\hat{V}_\alpha$ in the last line. This directly matches (\ref{eq:S_baby}), and we find that $V^*$ and $V^*_\alpha$ provide equivalent definitions of $\swap_\text{baby}$.

By (\ref{eq:V*exp}) for operators related by induced operator maps, we know that the expectation value of $\swap_\text{baby}$ must match those of both $\tilde{\swap}_\text{AdS}$ and $\tilde{\swap}_\alpha$ on states related to $|\psi_1\rangle$,
\begin{equation} \label{eq:exp_psi1}
    \langle \tilde{\swap}_\alpha \rangle_{V_\alpha|\psi_1\rangle} = \langle \tilde{\swap}_\text{AdS} \rangle_{V|\psi_1\rangle} = \langle \swap_\text{baby} \rangle_{|\psi_1\rangle}.
\end{equation}
Note that $\tilde\swap_\alpha$ and $\tilde\swap_\text{AdS}$ are common to both (\ref{eq:exp_psi2}) and (\ref{eq:exp_psi1}). If the AR condition (\ref{eq:ARcond}) holds, $V|\psi_1\rangle$ and $V_\text{HKLL}|\psi_2\rangle$ are the same state on $\hs_A$, and the expectation value of $\tilde\swap_\text{AdS}$ must be the same on both:
\begin{equation}
    \langle\tilde\swap_\text{AdS}\rangle_{V|\psi_1\rangle} = \langle\tilde\swap_\text{AdS}\rangle_{V_\text{HKLL}|\psi_2\rangle} = 1, \qquad \text{(AR condition holds).}
\end{equation}
In this case, (\ref{eq:exp_psi1}) also evaluates to 1 and is equivalent to (\ref{eq:exp_psi2}); this was verified for $\swap_\text{baby}$ in \cite{higginbotham_tests_2025}. However if the AR condition is broken, then (\ref{eq:exp_psi2}) and (\ref{eq:exp_psi1}) may differ.

All that remains is to define a SWAP test operator on $\hs_{A\beta'}$ for the $\beta$ observer. The only way to do this is by a reconstruction of $\swap_\text{baby}$, which we denote as $\tilde{\swap}_\beta$:
\begin{equation} \label{eq:S_beta}
    R^*_\beta : \swap_\text{baby} \to \tilde\swap_\beta
\end{equation}
Unfortunately, the non-isometry of $V_\beta$ prevents a natural definition of $R^*_\beta$. Therefore $R^*_\beta$ will generically be an approximate reconstruction that only preserves the expectation value up to some error,
\begin{equation} \label{eq:approx_expBeta}
    \langle \swap_\text{baby} \rangle_{|\psi_1\rangle} \approx \langle \tilde{\swap}_\beta \rangle_{V_\beta|\psi_1\rangle}.
\end{equation}
As a result, $\beta$'s prediction for the SWAP test will be limited by this error, as was found by EGH. However, this does not mean that a perfect prediction from $\beta$ is impossible -- there could be an operator in $\B(\hs_{A\beta'}\otimes\hs_{A\beta'})$ that is mapped directly to $\swap_\text{baby}$ by $V^*_\beta$,
\begin{equation} \label{eq:exactSbeta}
    V^\dagger_\beta \tilde{\swap}_\beta V_\beta = \swap_\text{baby}.
\end{equation}
If this is the case, $\tilde{\swap}_\beta$ would be an exact reconstruction whose expectation value precisely matches that of $\swap_\text{baby}$, and the description given by the $\beta$ observer would be exact.

All five SWAP test operators defined in this section -- and their relationships via reconstruction map $R^*$ or induced operator map $V^*$ -- are summarized in figure \ref{fig:Op_map}. Now that we have a general idea of how $\alpha$ and $\beta$ should perform the SWAP test, let us return to the specific observer rules to determine how each observer would describe the bulk. Following EGH, we begin with the HUZ rules.

\section{HUZ rules for observers} \label{sec:HUZ}

In \cite{engelhardt_observer_2025}, EGH used the HUZ rules \cite{harlow_quantum_2025} to determine how the $\alpha$ and $\beta$ observers would measure the SWAP test. We do the same in this section, applying the HUZ rules to the new SWAP test operators defined in section \ref{sec:swap}. We begin by reviewing the construction of the observer-modified holographic maps $\hat{V}_\alpha$, $V_\alpha$, and $V_\beta$ according to the HUZ rules.

Motivated by the assumption that observers are classical, the HUZ rules first apply a quantum-to-classical channel to the observer subsystem before applying a holographic map to the entire bulk state. This channel is implemented by cloning the observer to an external reference in a particular basis; the cloned state of the observer and its reference is denoted $|\omega\rangle$ in \cite{harlow_quantum_2025,engelhardt_observer_2025}. In this work, it will be helpful to have an operational notation for cloning. We draw inspiration from qubits, where cloning in the $Z$ basis can be done by applying a CNOT gate from the qubit to an ancilla $|0\rangle$. Therefore, will use a CNOT gate to denote the cloning operation preparing $|\omega\rangle$; for example, cloning the $\alpha$ observer to an external reference $\alpha'$ will be denoted as
\begin{equation}
    \vcenter{\hbox{
    \begin{tikzpicture}[thick]
        \draw (0,0) -- (0,1);
        \node[scale=0.9,anchor=north] at (0,0) {$|\psi_\alpha\rangle$};
        \draw (1,0) -- (1,1);
        \node[scale=0.9,anchor=north] at (1,0) {$|0\rangle_{\alpha'}$};
        \draw (0,0.5) -- (1.15,0.5);
        \draw[fill=black] (0,0.5) circle (0.1);
        \draw (1,0.5) circle (0.15);
        \node[anchor=east] at (-0.4,0.25) {$|\omega\rangle_{\alpha\alpha'} \equiv$};
    \end{tikzpicture}
    }}
\end{equation}
We do not intend to restrict cloning to be only in the $Z$ basis -- any basis can be chosen by conjugating CNOT with a unitary. Additionally, the CNOT gate can be generalized to perform cloning for $d$-dimensional qudits. This operational notation is only meant to make the role of cloning in later circuit diagrams more clear.

\begin{figure}
    \centering
    \begin{tikzpicture}[thick,scale=1.1]

\begin{scope}[shift={(0,0)},xscale=-1]
    \draw (0,0.25) -- (0,1);
    \node[anchor=north] at (0,0.25) {$M_a$};
    \draw (1,0.25) -- (1,1);
    \node[anchor=north] at (1,0.25) {$\alpha$};
    \draw (2,0.25) -- (2,2.75);
    \node[anchor=north,scale=0.8] at (2,0.25) {$|0\rangle_{\alpha'}$};
    \node[anchor=south] at (2,2.75) {$\alpha'$};
    \draw (1,0.625) -- (2.15,0.625);
    \draw[fill=black] (1,0.625) circle (0.1);
    \draw (2,0.625) circle (0.15);
    \draw[fill=gray!20] (-0.5,1) rectangle (1.5,2);
    \node[scale=1.2] at (0.5,1.5) {HKLL};
    \draw (0.5,2) -- (0.5,2.75);
    \node[anchor=south] at (0.5,2.75) {$A$};
    \node[anchor=west,scale=1.2] at (4.5, 1.5) {$\hat{V}_\alpha =$};
\end{scope}

\begin{scope}[shift={(0,-5)},xscale=-1]
    \draw (-2.5,0.25) -- (-2.5,1);
    \node[anchor=north] at (-2.5,0.25) {$\beta$};
    \draw (-1.5,0.25) -- (-1.5,1);
    \node[anchor=north] at (-1.5,0.25) {$M_b$};
    \draw[fill=gray!20] (-3,1) rectangle (-1,2);
    \node[scale=1.2] at (-2,1.5) {$O$};
    \draw (-2,2) -- (-2,2.75);
    \node[anchor=south] at (-2,2.75) {$\prescript{}{b}{\langle0|}$};
    \draw (0,0.25) -- (0,1);
    \node[anchor=north] at (0,0.25) {$M_a$};
    \draw (1,0.25) -- (1,1);
    \node[anchor=north] at (1,0.25) {$\alpha$};
    \draw (2,0.25) -- (2,2.75);
    \node[anchor=north,scale=0.8] at (2,0.25) {$|0\rangle_{\alpha'}$};
    \node[anchor=south] at (2,2.75) {$\alpha'$};
    \draw (1,0.625) -- (2.15,0.625);
    \draw[fill=black] (1,0.625) circle (0.1);
    \draw (2,0.625) circle (0.15);
    \draw[fill=gray!20] (-0.5,1) rectangle (1.5,2);
    \node[scale=1.2] at (0.5,1.5) {HKLL};
    \draw (0.5,2) -- (0.5,2.75);
    \node[anchor=south] at (0.5,2.75) {$A$};
    \node[anchor=west,scale=1.2] at (4.5, 1.5) {$V_\alpha = d_b^{1/2}$};
\end{scope}

\begin{scope}[shift={(0,-10)},xscale=-1]
    \draw (-2.5,0.25) -- (-2.5,1);
    \node[anchor=north] at (-2.5,0.25) {$\beta$};
    \draw (-1.5,0.25) -- (-1.5,1);
    \node[anchor=north] at (-1.5,0.25) {$M_b$};
    \draw[fill=gray!20] (-3,1) rectangle (-1,2);
    \node[scale=1.2] at (-2,1.5) {$O$};
    \draw (-2,2) -- (-2,2.75);
    \node[anchor=south] at (-2,2.75) {$\prescript{}{b}{\langle0|}$};
    \draw (-3.5,0.25) -- (-3.5,2.75);
    \node[anchor=north,scale=0.8] at (-3.5,0.25) {$|0\rangle_{\beta'}$};
    \node[anchor=south] at (-3.5,2.75) {$\beta'$};
    \draw (-2.5,0.625) -- (-3.65,0.625);
    \draw[fill=black] (-2.5,0.625) circle (0.1);
    \draw (-3.5,0.625) circle (0.15);
    \draw (0,0.25) -- (0,1);
    \node[anchor=north] at (0,0.25) {$M_a$};
    \draw (1,0.25) -- (1,1);
    \node[anchor=north] at (1,0.25) {$\alpha$};
    \draw[fill=gray!20] (-0.5,1) rectangle (1.5,2);
    \node[scale=1.2] at (0.5,1.5) {HKLL};
    \draw (0.5,2) -- (0.5,2.75);
    \node[anchor=south] at (0.5,2.75) {$A$};
    \node[anchor=west,scale=1.2] at (4.5, 1.5) {$V_\beta = d_b^{1/2}$};
\end{scope}

\end{tikzpicture}
    \caption{Circuit diagrams for the three observer-modified holographic maps constructed using the HUZ rules. A CNOT gate is used to denote the cloning operation, but the cloning can be taken to be in any basis of choice.}
    \label{fig:HUZ_rules}
\end{figure}
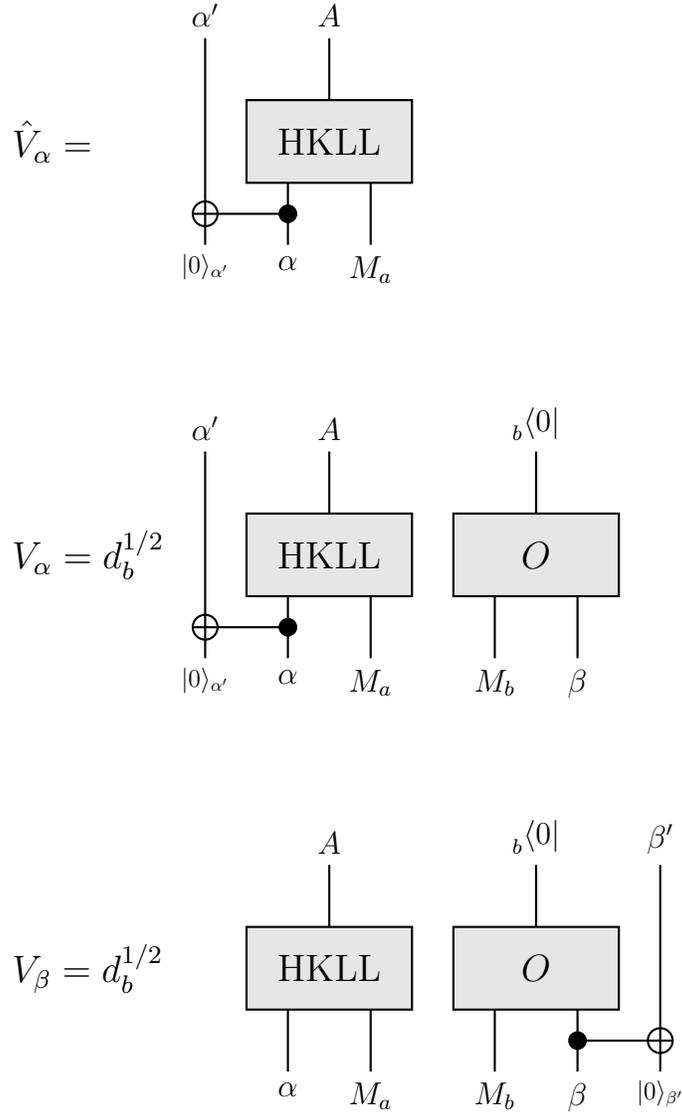

We now obtain constructions for the three observer-modified holographic maps by applying these rules to $V_\text{HKLL}$ and $V$ shown in figure \ref{fig:map2A}. $\hat{V}_\alpha$ is obtained by cloning $\alpha$ to $\alpha'$ in $V_\text{HKLL}$; $V_\alpha$ is obtained by cloning $\alpha$ to $\alpha'$ in $V$; and $V_\beta$ is obtained by cloning $\beta$ to $\beta'$ in $V$. Figure \ref{fig:HUZ_rules} shows circuit diagrams for all three maps. As alluded to in section \ref{sec:web}, we see that $\hat{V}_\alpha$ is isometric, while $V_\alpha$ and $V_\beta$ are non-isometric due to the post-selection on the baby universe.

\subsection{\texorpdfstring{$\alpha$}{alpha} observer} \label{sec:HUZ_alpha}
Let us now consider how the $\alpha$ observer would predict the result of the SWAP test. We found in section \ref{sec:swap} that $\alpha$ should measure $\tilde\swap_\alpha$ defined in (\ref{eq:S_alpha}). Using $\hat{V}_\alpha$ constructed from the HUZ rules as shown in figure \ref{fig:HUZ_rules}, we find
\begin{equation} \label{eq:S_alpha_circ}
    \tilde\swap_\alpha = 
    \vcenter{\hbox{
    \begin{tikzpicture}[thick,scale=0.8]

\newcommand{\CNOT}[4]{
    \draw[fill=black] (#1,#2) circle (0.1);
    \draw (#1,#2) -- (#3,#4);
    \draw[fill=white] (#3,#4) circle (0.15);
    \draw (#3+0.15,#4) -- (#3-0.15,#4);
    \draw (#3,#4+0.15) -- (#3,#4-0.15);
}

\begin{scope}[shift={(0,0)},xscale=-1,yscale=-1]
    \draw (0,0.25) -- (0,1);
    \node[scale=0.8,anchor=west] at (0,0.625) {$M_a$};
    \draw (1,0.25) -- (1,1);
    \node[scale=0.8,anchor=west] at (0.95,0.625) {$\alpha$};
    \draw (2,0.25) -- (2,2.75);
    \node[anchor=north] at (2,2.75) {$\alpha'$};
    \CNOT{1}{0.625}{2}{0.625};
    \draw[fill=gray!20] (-0.5,1) rectangle (1.5,2);
    \node at (0.5,1.5) {HKLL$^\dagger$};
    \draw (0.5,2) -- (0.5,2.75);
    \node[anchor=north] at (0.5,2.75) {$A$};
\end{scope}

\begin{scope}[shift={(2,0)},yscale=-1]
    \draw (0,0.25) -- (0,1);
    \node[scale=0.8,anchor=east] at (0,0.625) {$M_a$};
    \draw (1,0.25) -- (1,1);
    \node[scale=0.8,anchor=east] at (0.95,0.625) {$\alpha$};
    \draw (2,0.25) -- (2,2.75);
    \node[anchor=north] at (2,2.75) {$\alpha'$};
    \CNOT{1}{0.625}{2}{0.625};
    \draw[fill=gray!20] (-0.5,1) rectangle (1.5,2);
    \node at (0.5,1.5) {HKLL$^\dagger$};
    \draw (0.5,2) -- (0.5,2.75);
    \node[anchor=north] at (0.5,2.75) {$A$};
\end{scope}

\begin{scope}[shift={(0,0.5)},xscale=-1]
    \draw (0,0.25) -- (0,1);
    \draw (1,0.25) -- (1,1);
    \draw (2,0.25) -- (2,2.75);
    \node[anchor=south] at (2,2.75) {$\alpha'$};
    \CNOT{1}{0.625}{2}{0.625};
    \draw[fill=gray!20] (-0.5,1) rectangle (1.5,2);
    \node at (0.5,1.5) {HKLL};
    \draw (0.5,2) -- (0.5,2.75);
    \node[anchor=south] at (0.5,2.75) {$A$};
\end{scope}

\begin{scope}[shift={(2,0.5)}]
    \draw (0,0.25) -- (0,1);
    \draw (1,0.25) -- (1,1);
    \draw (2,0.25) -- (2,2.75);
    \node[anchor=south] at (2,2.75) {$\alpha'$};
    \CNOT{1}{0.625}{2}{0.625};
    \draw[fill=gray!20] (-0.5,1) rectangle (1.5,2);
    \node at (0.5,1.5) {HKLL};
    \draw (0.5,2) -- (0.5,2.75);
    \node[anchor=south] at (0.5,2.75) {$A$};
\end{scope}

\draw (0,-0.25) -- (2,0.75);
\draw (0,0.75) -- (2,-0.25);
\draw (-1,-0.25) -- (-1,0.75);
\draw (3,-0.25) -- (3,0.75);

\node[scale=0.8] at (-2,0.25) {$|0\rangle\langle0|_{\alpha'}$};
\node[scale=0.8] at (4,0.25) {$|0\rangle\langle0|_{\alpha'}$};

\draw[decorate,decoration={brace,amplitude=6pt,mirror}] (5,-0.25) -- (5,0.75);
\node[scale=1,anchor=west] at (5.25,0.25) {$\swap_\text{AdS}$};

\draw[decorate,decoration={brace,amplitude=6pt,mirror}] (5,0.8) -- (5,3.25);
\node[scale=1,anchor=west] at (5.25,2.025) {$\hat{V}_\alpha$};

\draw[decorate,decoration={brace,amplitude=6pt}] (5,-0.3) -- (5,-2.75);
\node[scale=1,anchor=west] at (5.25,-1.525) {$\hat{V}^\dagger_\alpha$};

\end{tikzpicture}
    }}
\end{equation}
where we have used $\text{CNOT}^\dagger = \text{CNOT}$. Comparing with EGH, who measure the SWAP test according to $\alpha$ using 
\begin{equation}
    \tilde\swap_\alpha^{(\text{EGH})} = \tilde\swap_\text{AdS} \otimes \id_{\alpha'},
\end{equation}
we find that our new definition of $\tilde\swap_\alpha$ differs in its action on the clone $\alpha'$. While $\tilde\swap_\alpha^{(\text{EGH})}$ acts simply as $\id_{\alpha'}$, $\tilde\swap_\alpha$ includes cloning operations and a projector onto the $|0\rangle$ subspace of $\alpha'$. 

Despite their differences, these operators actually act in the same way on the only two states of interest: $\hat{V}_\alpha|\psi_2\rangle$ \textit{without} a baby universe and $V_\alpha|\psi_1\rangle$ \textit{with} a baby universe. In both cases, the HKLL$^\dagger$ and CNOT$^\dagger$ operators in $\tilde\swap_\alpha$ always act to return $\alpha'$ to its initial $|0\rangle$ state. As a result, the projector will have no effect and can be replaced with the identity.\footnote{This is reminiscent of the dynamically generated subspace defined in \cite{dewolfe_nonisometric_2023,dewolfe_bulk_2024} for black hole non-isometric codes.} Since $\text{CNOT}^2=\id$, the cloning operators also drop out and $\tilde\swap_\alpha$ reduces to
\begin{equation}
    \tilde\swap_\alpha = \tilde\swap_\alpha^{(\text{EGH})} \qquad \text{on } \hat{V}_\alpha|\psi_2\rangle \text{ and } V_\alpha|\psi_1\rangle.
\end{equation}
Therefore EGH's choice is correct on these states, and we will continue with their definition. 

In appendix A of \cite{engelhardt_observer_2025}, EGH finds the expectation values of this SWAP test operator in these two states to be 
\begin{align}
    \langle\tilde{\swap}_\text{AdS}\otimes\id_{\alpha'}\rangle_{\hat{V}_\alpha |\psi_2\rangle} &= 1 \label{eq:EGHeq7} \\
    \int dO \, \langle \tilde{\swap}_\text{AdS} \otimes \mathbb{1}_{\alpha'} \rangle_{V_\alpha|\psi_1\rangle} &= \frac{d_b}{d_b + 2} \Big[ 1 + \tr \big( \psi_{1,M_a}^2 \big) + \tr \big( \psi_{1,M_a} \psi_{1,M_a}^T \big) \Big] \label{eq:dO_exp1}
\end{align}
where $\psi_{1,M_a} \equiv \tr_{M_b} ( |\psi_1\rangle\langle\psi_1|_{M_a M_b} )$ and the orthogonal matrix $O$ appearing in $V_\alpha$ has been averaged over the Haar measure. Equation (\ref{eq:EGHeq7}) matches the expectation value of the bulk SWAP test $\swap_\text{AdS}$ given by (\ref{eq:expSads}), confirming that the $\alpha$ observer sees $M_a$ in a pure state $|\psi_2\rangle$. Since $|\psi_1\rangle$ is generically entangled between $M_a$ and $M_b$, (\ref{eq:dO_exp1}) does not match this result, even in the large $d_b$ limit. Therefore, EGH concludes that $\alpha$ can only describe the bulk state as $|\psi_2\rangle$ without a baby universe.

However, we found in section \ref{sec:newops} that if both equation (\ref{eq:exp_psi1}) and the AR condition (\ref{eq:ARcond}) hold, then $\langle\tilde\swap_\alpha\rangle_{V_\alpha|\psi_1\rangle} = 1$. Since (\ref{eq:dO_exp1}) is in conflict with this, one of these two assumptions must be broken. Let us first check that (\ref{eq:exp_psi1}) holds; this requires that the two definitions (\ref{eq:S_baby}) and (\ref{eq:S_baby2}) for $\swap_\text{baby}$ are equivalent, which is true if $V^\dagger_\alpha \hat{V}_\alpha = V^\dagger V_\text{HKLL}$. Indeed this holds for the HUZ rules:
\begin{equation}
    V^\dagger_\alpha \hat{V}_\alpha =
    \vcenter{\hbox{\begin{tikzpicture}[thick,scale=0.8]

\newcommand{\CNOT}[4]{
    \draw[fill=black] (#1,#2) circle (0.1);
    \draw (#1,#2) -- (#3,#4);
    \draw[fill=white] (#3,#4) circle (0.15);
    \draw (#3+0.15,#4) -- (#3-0.15,#4);
    \draw (#3,#4+0.15) -- (#3,#4-0.15);
}

\draw (0,0.25) -- (0,1);
\node[anchor=north] at (0,0.25) {$M_a$};
\draw (1,0.25) -- (1,1);
\node[anchor=north] at (1,0.25) {$\alpha$};
\draw (2,0.25) -- (2,2.75);
\node[scale=0.8,anchor=north] at (2,0.25) {$|0\rangle_{\alpha'}$};
\CNOT{1}{0.625}{2}{0.625};
\draw[fill=gray!20] (-0.5,1) rectangle (1.5,2);
\node at (0.5,1.5) {HKLL};
\draw (0.5,2) -- (0.5,2.75);

\begin{scope}[shift={(0,4.75)},yscale=-1]
    \draw (-2.5,0.25) -- (-2.5,1);
    \node[anchor=south] at (-2.5,0.25) {$\beta$};
    \draw (-1.5,0.25) -- (-1.5,1);
    \node[anchor=south] at (-1.5,0.25) {$M_b$};
    \draw[fill=gray!20] (-3,1) rectangle (-1,2);
    \node at (-2,1.5) {$O^T$};
    \draw (-2,2) -- (-2,2.75);
    \node[anchor=north] at (-2,2.75) {$|0\rangle_b$};
    \draw (0,0.25) -- (0,1);
    \node[anchor=south] at (0,0.25) {$M_a$};
    \draw (1,0.25) -- (1,1);
    \node[anchor=south] at (1,0.25) {$\alpha$};
    \draw (2,0.25) -- (2,2);
    \node[scale=0.8,anchor=south] at (2,0.25) {$\prescript{}{\alpha'}{\langle0|}$};
    \CNOT{1}{0.625}{2}{0.625};
    \draw[fill=gray!20] (-0.5,1) rectangle (1.5,2);
    \node at (0.5,1.5) {HKLL$^\dagger$};
\end{scope}

\end{tikzpicture}}}
    = O^T |0\rangle_b \otimes \id_{M_a \alpha} = 
    \vcenter{\hbox{\begin{tikzpicture}[thick,scale=0.8]

\draw (0,0.25) -- (0,1);
\node[anchor=north] at (0,0.25) {$M_a$};
\draw (1,0.25) -- (1,1);
\node[anchor=north] at (1,0.25) {$\alpha$};
\draw[fill=gray!20] (-0.5,1) rectangle (1.5,2);
\node at (0.5,1.5) {HKLL};
\draw (0.5,2) -- (0.5,2.75);

\begin{scope}[shift={(0,4.75)},yscale=-1]
    \draw (-2.5,0.25) -- (-2.5,1);
    \node[anchor=south] at (-2.5,0.25) {$\beta$};
    \draw (-1.5,0.25) -- (-1.5,1);
    \node[anchor=south] at (-1.5,0.25) {$M_b$};
    \draw[fill=gray!20] (-3,1) rectangle (-1,2);
    \node at (-2,1.5) {$O^T$};
    \draw (-2,2) -- (-2,2.75);
    \node[anchor=north] at (-2,2.75) {$|0\rangle_b$};
    \draw (0,0.25) -- (0,1);
    \node[anchor=south] at (0,0.25) {$M_a$};
    \draw (1,0.25) -- (1,1);
    \node[anchor=south] at (1,0.25) {$\alpha$};
    \draw[fill=gray!20] (-0.5,1) rectangle (1.5,2);
    \node at (0.5,1.5) {HKLL$^\dagger$};
\end{scope}

\end{tikzpicture}}}
    = V^\dagger V_\text{HKLL}
\end{equation}
Therefore (\ref{eq:exp_psi1}) holds and the disagreement between (\ref{eq:EGHeq7}) and (\ref{eq:dO_exp1}) must indicate that
\begin{equation}
    \langle \tilde{\swap}_\text{AdS} \rangle_{V_\text{HKLL}|\psi_2\rangle} \neq \langle \tilde{\swap}_\text{AdS} \rangle_{V|\psi_1\rangle}.
\end{equation}
This can only be true if $V_\text{HKLL}|\psi_2\rangle$ and $V|\psi_1\rangle$ are different states on $\hs_{A}$. Somehow, (\ref{eq:dO_exp1}) must have implicitly violated the AR condition (\ref{eq:ARcond})! To identify the source of this violation, let us rewrite the AR condition using density matrices $\psi_i \equiv |\psi_i\rangle\langle\psi_i|$ ($i=1,2$) and average over the orthogonal $O$ in $V$,
\begin{equation} \label{eq:ARcond_dens}
    \int dO \, V\psi_1V^\dagger = V_\text{HKLL} \psi_2 V^\dagger_\text{HKLL}.
\end{equation}
Performing the average, we find the left hand side reduces to
\begin{equation} \label{eq:ARcond_dens_left}
    \int dO \, d_b \,
    \vcenter{\hbox{
    \begin{tikzpicture}[thick,scale=0.8]

\draw[fill=gray!20] (2.5,0.5) rectangle (1.5,-0.5);
\node[scale=1.2] at (2,0) {$\psi_\alpha$};
\draw[fill=gray!20] (-1,-0.5) rectangle (1,0.5);
\node[scale=1.2] at (0,0) {$\psi_1$};
\draw[fill=gray!20] (-1.5,-0.5) rectangle (-2.5,0.5);
\node[scale=1.2] at (-2,0) {$\psi_\beta$};

\begin{scope}[shift={(0.5,0.25)}]
    \draw (-2.5,0.25) -- (-2.5,1);
    \draw (-1,0.25) -- (-1,1);
    \node[anchor=east,scale=0.8] at (-1,0.625) {$M_b$};
    \draw[fill=gray!20] (-2.75,1) rectangle (-0.75,2);
    \node[scale=1] at (-1.75,1.5) {$O$};
    \draw (-1.75,2) -- (-1.75,2.75);
    \node[anchor=south] at (-1.75,2.75) {$\prescript{}{b}{\langle0|}$};
    \draw (0,0.25) -- (0,1);
    \node[anchor=west,scale=0.8] at (0,0.625) {$M_a$};
    \draw (1.5,0.25) -- (1.5,1);
    \draw[fill=gray!20] (-0.25,1) rectangle (1.75,2);
    \node[scale=1] at (0.75,1.5) {HKLL};
    \draw (0.75,2) -- (0.75,2.75);
    \node[anchor=south] at (0.75,2.75) {$A$};
\end{scope}

\begin{scope}[shift={(0.5,-0.25)},yscale=-1]
    \draw (-2.5,0.25) -- (-2.5,1);
    \draw (-1,0.25) -- (-1,1);
    \draw[fill=gray!20] (-2.75,1) rectangle (-0.75,2);
    \node[scale=1] at (-1.75,1.5) {$O^T$};
    \draw (-1.75,2) -- (-1.75,2.75);
    \node[anchor=north] at (-1.75,2.75) {$|0\rangle_b$};
    \draw (0,0.25) -- (0,1);
    \draw (1.5,0.25) -- (1.5,1);
    \draw[fill=gray!20] (-0.25,1) rectangle (1.75,2);
    \node[scale=1] at (0.75,1.5) {HKLL};
    \draw (0.75,2) -- (0.75,2.75);
    \node[anchor=north] at (0.75,2.75) {$A$};
\end{scope}

\end{tikzpicture}
    }}
    =
    \vcenter{\hbox{
    \begin{tikzpicture}[thick,scale=0.8]

\draw[fill=gray!20] (2.5,0.5) rectangle (1.5,-0.5);
\node[scale=1.2] at (2,0) {$\psi_\alpha$};
\draw[fill=gray!20] (-1,-0.5) rectangle (1,0.5);
\node[scale=1.2] at (0,0) {$\psi_1$};

\begin{scope}[shift={(0.5,0.25)}]
    \draw (-1,0.25) -- (-1,1) -- (-2,1) -- (-2,-0.25);
    \node[anchor=east,scale=0.8] at (-1,0.625) {$M_b$};
    \draw (0,0.25) -- (0,1);
    \node[anchor=west,scale=0.8] at (0,0.625) {$M_a$};
    \draw (1.5,0.25) -- (1.5,1);
    \draw[fill=gray!20] (-0.25,1) rectangle (1.75,2);
    \node[scale=1] at (0.75,1.5) {HKLL};
    \draw (0.75,2) -- (0.75,2.75);
    \node[anchor=south] at (0.75,2.75) {$A$};
\end{scope}

\begin{scope}[shift={(0.5,-0.25)},yscale=-1]
    \draw (-1,0.25) -- (-1,1) -- (-2,1) -- (-2,-0.25);
    \draw (0,0.25) -- (0,1);
    \draw (1.5,0.25) -- (1.5,1);
    \draw[fill=gray!20] (-0.25,1) rectangle (1.75,2);
    \node[scale=1] at (0.75,1.5) {HKLL};
    \draw (0.75,2) -- (0.75,2.75);
    \node[anchor=north] at (0.75,2.75) {$A$};
\end{scope}

\end{tikzpicture}
    }}
\end{equation}
Equating the right hand sides of (\ref{eq:ARcond_dens}) and (\ref{eq:ARcond_dens_left}), we find the AR condition can only be satisfied for average $O$ if
\begin{equation}
    \tr_{M_b} \big( |\psi_1\rangle\langle\psi_1|_{M_a M_b} \big) = |\psi_2\rangle \langle\psi_2|_{M_a}.
\end{equation}
This is inconsistent with our assumption that $M_a$ is entangled in $|\psi_1\rangle$ and pure in $|\psi_2\rangle$. Therefore, there is no choice of $|\psi_1\rangle$ and $|\psi_2\rangle$ which satisfies the AR condition on average.\footnote{It is possible to preserve the AR condition on average by entangling $|\psi_1\rangle_{M_a,M_b}$ and $|\psi_2\rangle_{M_a}$ with an external reference. This can be done in the AS$^2$ construction by entangling the external reference with the heavy operator preparing the CFT states. We thank Luca Ciambelli and Beni Yoshida for discussions on this point. We refer the reader to \cite{higginbotham_tests_2025,mori_baby_2025} for further details on such a setup.} This is the source of the disagreement between (\ref{eq:EGHeq7}) and (\ref{eq:dO_exp1}): the average in the latter always picks up contributions from choices of $O$ that violate the AR condition.

\paragraph{Satisfying AR in a toy qubit model.} Instead of averaging over $O$, let us choose a particular $O$ so that the AR condition is satisfied. In this case, we expect $\langle \tilde\swap_\alpha \rangle_{V_\alpha |\psi_1\rangle}$ to be 1. To construct an example of this, let us consider a toy qubit model for the two bulk states. We will model the entanglement between $M_a$ and $M_b$ by taking $|\psi_1\rangle_{M_a M_b}$ to be given by a single maximally entangled pair of qubits:
\begin{equation} \label{eq:Phi}
    |\psi_1\rangle_{M_a M_b} = \frac{1}{\sqrt{2}} \big( |0\rangle_{M_a} |0\rangle_{M_b} + |1\rangle_{M_a} |1\rangle_{M_b} \big) \equiv |\Phi^+\rangle_{M_a M_b}.
\end{equation}
For simplicity, we will also take $\beta$ to be given by a single qubit. Let us consider a choice of $O$ such that we post-select on a similar maximally entangled state between $\beta$ and $M_b$,
\begin{equation}
    \prescript{}{b}{\langle0|O} = \prescript{}{\beta M_b}{\langle\Phi^+|}. 
\end{equation}
With this choice of matter state and post-selection, both $V$ and $V_\alpha$ act to teleport the state of $\beta$ to $M_a$ where it is then encoded in $A$ by the HKLL map:
\begin{equation}
    V_\alpha \big( |\Phi^+\rangle_{M_a M_b} |\psi_\alpha\rangle |\psi_\beta\rangle \big) = \,2
    \vcenter{\hbox{
    \begin{tikzpicture}[thick,scale=0.8]

\newcommand{\CNOT}[4]{
    \draw[fill=black] (#1,#2) circle (0.1);
    \draw (#1,#2) -- (#3,#4);
    \draw[fill=white] (#3,#4) circle (0.15);
    \draw (#3+0.15,#4) -- (#3-0.15,#4);
    \draw (#3,#4+0.15) -- (#3,#4-0.15);
}

\draw (0,0) -- (0,2.5) -- (0.5,2.75) -- (1,2.5) -- (1,0) -- (1.5,-0.25) -- (2,0) -- (2,0.75);
\node[scale=0.9,anchor=north] at (0,0) {$|\psi_\beta\rangle$};
\node[scale=0.9,anchor=north] at (1,0) {$M_b$};
\node[scale=0.9,anchor=north] at (2,0) {$M_a$};

\draw (3,0) -- (3,0.75);
\node[scale=0.9,anchor=north] at (3,0) {$|\psi_\alpha\rangle$};

\draw (4,0) -- (4,2.5);
\node[scale=0.9,anchor=north] at (4,0) {$|0\rangle_{\alpha'}$};

\CNOT{3}{0.375}{4}{0.375};

\draw[fill=gray!20] (1.5,0.75) rectangle (3.5,1.75);
\node[scale=1] at (2.5,1.25) {HKLL};

\draw (2.5,1.75) -- (2.5,2.5);
\node[scale=0.9,anchor=south] at (2.5,2.5) {$A$};


\end{tikzpicture}
    }}
\end{equation}
Therefore, no pure state on $\beta$ is in the kernel of $V$, and the AR condition is satisfied if $|\psi_2\rangle_{M_a} = |\psi_\beta\rangle$. In this case, the expectation value of $\tilde\swap_\alpha = \tilde\swap_\text{AdS}\otimes \id_{\alpha'}$ is given by
\begin{equation}
    \langle \tilde{\swap}_\text{AdS}\otimes\id_{\alpha'} \rangle_{V_\alpha |\Phi^+\rangle} = \,16
    \vcenter{\hbox{
    \begin{tikzpicture}[thick,scale=0.8]

\draw (0,0) -- (0,0.75) -- (0.5,1) -- (1,0.75) -- (1,0) -- (1.5,-0.25) -- (2,0) -- (2,0.75) -- (3,1.5) -- (3,2.25) -- (3.5,2.5) -- (4,2.25) -- (4,1.5) -- (4.5,1.25) -- (5,1.5) -- (5,2.25);

\draw (0,2.25) -- (0,1.5) -- (0.5,1.25) -- (1,1.5) -- (1,2.25) -- (1.5,2.5) -- (2,2.25) -- (2,1.5) -- (3,0.75) -- (3,0) -- (3.5,-0.25) -- (4,0) -- (4,0.75) -- (4.5,1) -- (5,0.75) -- (5,0);

\node[scale=0.8,anchor=north] at (0,0) {$|\psi_\beta\rangle$};
\node[scale=0.8,anchor=north] at (5,0) {$|\psi_\beta\rangle$};

\node[scale=0.8,anchor=south] at (0,2.25) {$\langle\psi_\beta|$};
\node[scale=0.8,anchor=south] at (5,2.25) {$\langle\psi_\beta|$};

\node[scale=0.8,anchor=north] at (1,0) {$M_b$};
\node[scale=0.8,anchor=north] at (4,0) {$M_b$};

\node[scale=0.8,anchor=north] at (2,0) {$M_a$};
\node[scale=0.8,anchor=north] at (3,0) {$M_a$};


\end{tikzpicture}
    }}
    =1,
\end{equation}
where the HKLL and cloning operations have been removed due to their isometry. In this toy model, the expectation value of $\tilde\swap_\alpha = \tilde{\swap}_\text{AdS}\otimes\id_{\alpha'}$ in $V_\alpha |\psi_1\rangle$ is 1, as expected.

\paragraph{How does $\alpha$ describe the bulk?} When the AR condition (\ref{eq:ARcond}) is satisfied, as in the above toy model, $V|\psi_1\rangle$ and $V_\text{HKLL}|\psi_2\rangle$ are the same state in $\hs_A$, and $\tilde\swap_\text{AdS}$ will not be able to distinguish them. In the same way, $V_\alpha|\psi_1\rangle$ and $\hat{V}_\alpha|\psi_2\rangle$ are the same state in the $\alpha$ observer's fundamental Hilbert space $\hs_{A\alpha'}$, and $\tilde\swap_\alpha$ will not be able to distinguish them. Therefore, $\alpha$ can equally well describe the bulk using $|\psi_1\rangle$ with a baby universe or $|\psi_2\rangle$ without one.

When the AR condition is broken, as is the case for average $O$ leading to (\ref{eq:EGHeq7}) and (\ref{eq:dO_exp1}), a measurement of $\tilde{\swap}_\alpha = \tilde{\swap}_\text{AdS} \otimes \mathbb{1}_{\alpha'}$ will be able to distinguish the two bulk states $|\psi_1\rangle$ and $|\psi_2\rangle$. We know from (\ref{eq:exp_psi2}) and (\ref{eq:exp_psi1}) that the $\alpha$ observer can be confident that their answer will agree with $\tilde{\swap}_\text{AdS}$ on $\hs_A$, no matter the result. Therefore we do not find either bulk state more valid than the other, only that they are distinguishable by $\alpha$ in this case.

\subsection{\texorpdfstring{$\beta$}{beta} observer} \label{sec:HUZ_beta}
Let us now consider how the $\beta$ observer would predict the measurement of the SWAP test. We found in section \ref{sec:swap} that $\beta$ should measure $\tilde\swap_\beta$ defined by the reconstruction of $\swap_\text{baby}$ using $R^*_\beta$, but the non-isometry of $V_\beta$ in figure \ref{fig:HUZ_rules} generically prevents a perfect reconstruction. Following (\ref{eq:approx_expBeta}), we will quantify the error in $\beta$'s prediction for the SWAP test as
\begin{equation} \label{eq:error}
    \Delta \equiv \langle \swap_\text{baby} \rangle_{|\psi_1\rangle} - \langle \tilde\swap_\beta \rangle_{V_\beta|\psi_1\rangle}.
\end{equation}
It will be useful to compute $\langle \swap_\text{baby} \rangle_{|\psi_1\rangle}$ for average choices of the orthogonal $O$ appearing in the definition (\ref{eq:S_baby}) of $\swap_\text{baby}$:
\begin{align*}
    \int dO\, \langle \swap_\text{baby} \rangle_{|\psi_1\rangle} &= \int dO\, d_b\, \tr \Big[ \big(\psi_1 \otimes \psi_\alpha \otimes \psi_\beta \big)^{\otimes2} \big( \swap_\text{AdS} \otimes O^T |0\rangle\langle0| O \big) \Big] \\
        &= \frac{d_b}{d_b + 2} \Big[ 1 + \tr\big( \psi_{1,M_a}^2 \big) + \tr\big( \psi_\beta \psi_\beta^T \big) \tr \big( \psi_1 \psi_1^{T_{M_b}} \big) \Big]. \numberthis \label{eq:expSbaby}
\end{align*}
Note that this is not 1 since the average over $O$ picks up contributions that violate the AR condition. Fortunately, we do not need to worry about preserving the AR condition here since (\ref{eq:approx_expBeta}) only depends on $|\psi_1\rangle$.

But how should we choose to define the reconstruction $\tilde\swap_\beta$? Similar to their treatment of the $\alpha$ observer, EGH chooses to measure the SWAP test according to $\beta$ using $\tilde\swap_\text{AdS}\otimes\id_{\beta'}$, and we will interpret this operator as a candidate reconstruction for $\tilde\swap_\beta$. In appendix A of \cite{engelhardt_observer_2025}, EGH finds the expectation value of this operator to be 
\begin{equation} \label{eq:EGH8}
    \int dO \, \langle \tilde{\swap}_\text{AdS} \otimes \mathbb{1}_{\beta'} \rangle_{V_\beta |\psi_1\rangle} = \frac{d_b}{d_b + 2} \Big[ \tr \big( \omega_\beta^2 \big) + \tr \big( \psi_{1,M_a}^2 \big) + \tr \big( \omega_\beta \omega_\beta^T \big) \tr \big( \psi_1 \psi_1^{T_{M_b}} \big) \Big]
\end{equation}
where the average has been taken over the orthogonal $O$ appearing in $V_\beta$, $\omega_\beta$ is the state of the $\beta$ observer after its clone has been traced out, and $T_{M_b}$ denotes a partial transpose on the $M_b$ subsystem of $\psi_1$. Using (\ref{eq:error}) and (\ref{eq:expSbaby}), we find the error in this choice of reconstruction is
\begin{align*} 
    \overline{\Delta}_\id &= \frac{d_b}{d_b + 2} \Big[ 1 - \tr \big( \omega_\beta^2 \big) + \Big( \tr \big( \psi_\beta \psi_\beta^T \big) - \tr \big( \omega_\beta\omega_\beta^T \big) \Big) \tr \big( \psi_1 \psi_1^{T_{M_b}} \big) \Big] \\
        &\approx 1 - \tr \big(\omega_\beta^2\big) \numberthis \label{eq:delta1}
\end{align*}
where a bar has been used to emphasize that this error was calculated for average $O$. In the second line, we have taken the large $d_b$ limit and dropped subleading terms involving transposes. This way of measuring the error in $\beta$'s description is consistent with EGH's results, who also found errors on the order of $\tr (\omega_\beta^2) = e^{-S_2(\omega_\beta)}$.

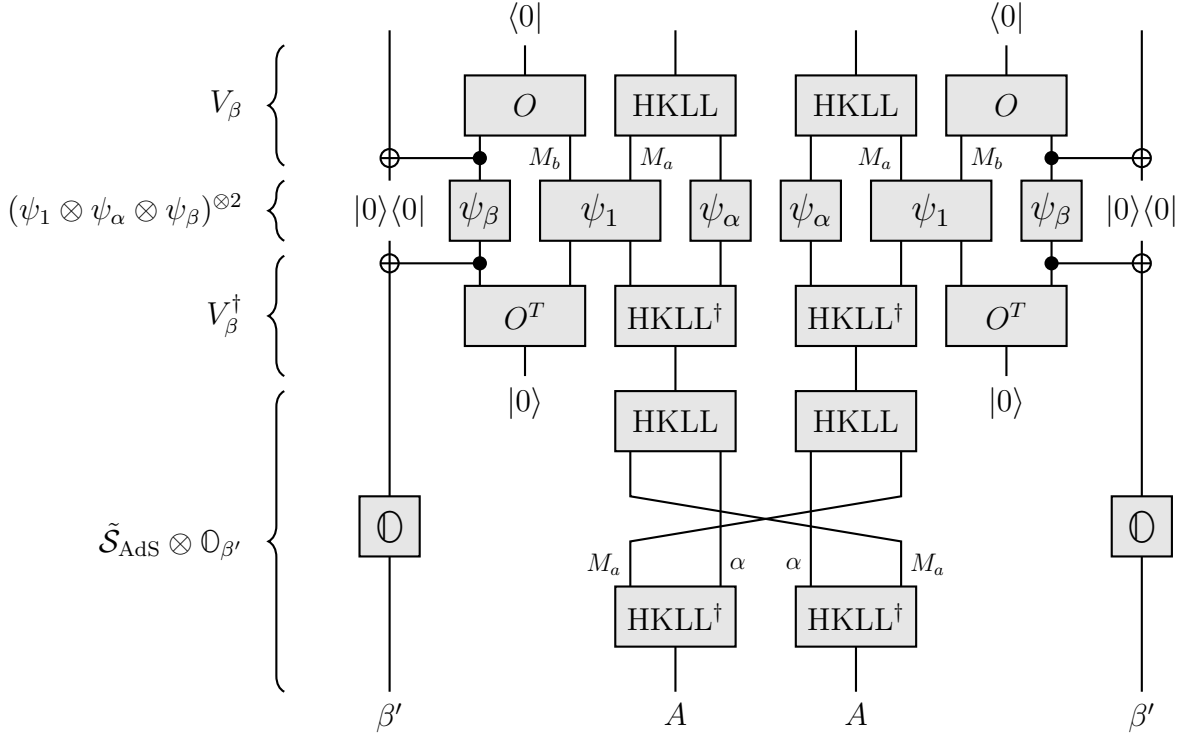
\begin{figure}[t]
    \centering
    \begin{tikzpicture}[thick, scale=0.8]

\newcommand{\CNOT}[4]{
    \draw[fill=black] (#1,#2) circle (0.1);
    \draw (#1,#2) -- (#3,#4);
    \draw[fill=white] (#3,#4) circle (0.15);
    \draw (#3+0.15,#4) -- (#3-0.15,#4);
    \draw (#3,#4+0.15) -- (#3,#4-0.15);
}


\draw[fill=gray!20] (0.25,-0.5) rectangle (1.25,0.5);
\node[scale=1.2] at (0.75,0) {$\psi_\alpha$};
\draw[fill=gray!20] (1.75,-0.5) rectangle (3.75,0.5);
\node[scale=1.2] at (2.75,0) {$\psi_1$};
\draw[fill=gray!20] (4.25,-0.5) rectangle (5.25,0.5);
\node[scale=1.2] at (4.75,0) {$\psi_\beta$};
\node[scale=1] at (6.25,0) {$|0\rangle\langle0|$};

\begin{scope}[xscale=-1]
\draw[fill=gray!20] (0.25,-0.5) rectangle (1.25,0.5);
\node[scale=1.2] at (0.75,0) {$\psi_\alpha$};
\draw[fill=gray!20] (1.75,-0.5) rectangle (3.75,0.5);
\node[scale=1.2] at (2.75,0) {$\psi_1$};
\draw[fill=gray!20] (4.25,-0.5) rectangle (5.25,0.5);
\node[scale=1.2] at (4.75,0) {$\psi_\beta$};
\node[scale=1] at (6.25,0) {$|0\rangle\langle0|$};
\end{scope}

\begin{scope}[yscale=-1]


\draw (0.75,0.5) -- (0.75,1.25);
\draw (2.25,0.5) -- (2.25,1.25);
\draw (3.25,0.5) -- (3.25,1.25);
\draw (4.75,0.5) -- (4.75,1.25);
\draw (6.25,0.5) -- (6.25,4.75);
\CNOT{4.75}{0.875}{6.25}{0.875};
\draw[fill=gray!20] (0.5,1.25) rectangle (2.5,2.25);
\node[scale=1] at (1.5,1.75) {HKLL$^\dagger$};
\draw (1.5,2.25) -- (1.5,3);
\draw[fill=gray!20] (3,1.25) rectangle (5,2.25);
\node[scale=1] at (4,1.75) {$O^T$};
\draw (4,2.25) -- (4,2.75);
\node[scale=1,anchor=north] at (4,2.75) {$|0\rangle$};
\draw[fill=gray!20] (0.5,3) rectangle (2.5,4);
\node[scale=1] at (1.5,3.5) {HKLL};
\draw (0.75,4) -- (0.75,4.75) -- (0.75,6.25);
\node[scale=0.8,anchor=east] at (0.75,5.875) {$\alpha$};
\draw (2.25,4) -- (2.25,4.75) -- (-2.25,5.5) -- (-2.25,6.25);
\node[scale=0.8,anchor=west] at (2.25,5.875) {$M_a$};
\draw[fill=gray!20] (0.5,6.25) rectangle (2.5,7.25);
\node[scale=1] at (1.5,6.75) {HKLL$^\dagger$};
\draw (1.5,7.25) -- (1.5,8);
\node[scale=1,anchor=north] at (1.5,8) {$A$};
\draw[fill=gray!20] (5.75,4.75) rectangle (6.75,5.75);
\node[scale=1.4] at (6.25,5.25) {$\mathbb{O}$};
\draw (6.25,5.75) -- (6.25,8);
\node[scale=1,anchor=north] at (6.25,8) {$\beta'$};


\begin{scope}[xscale=-1]
\draw (0.75,0.5) -- (0.75,1.25);
\draw (2.25,0.5) -- (2.25,1.25);
\draw (3.25,0.5) -- (3.25,1.25);
\draw (4.75,0.5) -- (4.75,1.25);
\draw (6.25,0.5) -- (6.25,4.75);
\CNOT{4.75}{0.875}{6.25}{0.875};
\draw[fill=gray!20] (0.5,1.25) rectangle (2.5,2.25);
\node[scale=1] at (1.5,1.75) {HKLL$^\dagger$};
\draw (1.5,2.25) -- (1.5,3);
\draw[fill=gray!20] (3,1.25) rectangle (5,2.25);
\node[scale=1] at (4,1.75) {$O^T$};
\draw (4,2.25) -- (4,2.75);
\node[scale=1,anchor=north] at (4,2.75) {$|0\rangle$};
\draw[fill=gray!20] (0.5,3) rectangle (2.5,4);
\node[scale=1] at (1.5,3.5) {HKLL};
\draw (0.75,4) -- (0.75,4.75) -- (0.75,6.25);
\node[scale=0.8,anchor=west] at (0.75,5.875) {$\alpha$};
\draw (2.25,4) -- (2.25,4.75) -- (-2.25,5.5) -- (-2.25,6.25);
\node[scale=0.8,anchor=east] at (2.25,5.875) {$M_a$};
\draw[fill=gray!20] (0.5,6.25) rectangle (2.5,7.25);
\node[scale=1] at (1.5,6.75) {HKLL$^\dagger$};
\draw (1.5,7.25) -- (1.5,8);
\node[scale=1,anchor=north] at (1.5,8) {$A$};
\draw[fill=gray!20] (5.75,4.75) rectangle (6.75,5.75);
\node[scale=1.4] at (6.25,5.25) {$\mathbb{O}$};
\draw (6.25,5.75) -- (6.25,8);
\node[scale=1,anchor=north] at (6.25,8) {$\beta'$};
\end{scope}

\end{scope}

\begin{scope}[yscale=1]


\draw (0.75,0.5) -- (0.75,1.25);
\draw (2.25,0.5) -- (2.25,1.25);
\node[scale=0.8,anchor=east] at (2.25,0.875) {$M_a$};
\draw (3.25,0.5) -- (3.25,1.25);
\node[scale=0.8,anchor=west] at (3.25,0.875) {$M_b$};
\draw (4.75,0.5) -- (4.75,1.25);
\draw (6.25,0.5) -- (6.25,3);
\CNOT{4.75}{0.875}{6.25}{0.875};
\draw[fill=gray!20] (0.5,1.25) rectangle (2.5,2.25);
\node[scale=1] at (1.5,1.75) {HKLL};
\draw (1.5,2.25) -- (1.5,3);
\draw[fill=gray!20] (3,1.25) rectangle (5,2.25);
\node[scale=1] at (4,1.75) {$O$};
\draw (4,2.25) -- (4,2.75);
\node[scale=1,anchor=south] at (4,2.75) {$\langle0|$};


\begin{scope}[xscale=-1]
\draw (0.75,0.5) -- (0.75,1.25);
\draw (2.25,0.5) -- (2.25,1.25);
\node[scale=0.8,anchor=west] at (2.25,0.875) {$M_a$};
\draw (3.25,0.5) -- (3.25,1.25);
\node[scale=0.8,anchor=east] at (3.25,0.875) {$M_b$};
\draw (4.75,0.5) -- (4.75,1.25);
\draw (6.25,0.5) -- (6.25,3);
\CNOT{4.75}{0.875}{6.25}{0.875};
\draw[fill=gray!20] (0.5,1.25) rectangle (2.5,2.25);
\node[scale=1] at (1.5,1.75) {HKLL};
\draw (1.5,2.25) -- (1.5,3);
\draw[fill=gray!20] (3,1.25) rectangle (5,2.25);
\node[scale=1] at (4,1.75) {$O$};
\draw (4,2.25) -- (4,2.75);
\node[scale=1,anchor=south] at (4,2.75) {$\langle0|$};
\end{scope}

\end{scope}


\draw[decorate,decoration={brace,amplitude=6pt}] (-8,-0.5) -- (-8,0.5);
\node[scale=1,anchor=east] at (-8.5,0) {$(\psi_1\otimes\psi_\alpha\otimes\psi_\beta)^{\otimes2}$};

\draw[decorate,decoration={brace,amplitude=6pt}] (-8,0.75) -- (-8,2.75);
\node[scale=1,anchor=east] at (-8.5,1.75) {$V_\beta$};

\draw[decorate,decoration={brace,amplitude=6pt}] (-8,-2.75) -- (-8,-0.75);
\node[scale=1,anchor=east] at (-8.5,-1.75) {$V_\beta^\dagger$};

\draw[decorate,decoration={brace,amplitude=6pt}] (-8,-8) -- (-8,-3);
\node[scale=1,anchor=east] at (-8.5,-5.5) {$\tilde\swap_\text{AdS}\otimes\mathbb{O}_{\beta'}$};

\end{tikzpicture}
    \caption{A circuit diagram depiction for $V_\beta (\psi_1 \otimes \psi_\alpha \otimes \psi_\beta)^{\otimes2} V^\dagger_\beta \big( \tilde{\swap}_\text{AdS} \otimes \mathbb{O}_{\beta'} \big)$ appearing in the first line of (\ref{eq:expRbaby}). Note that all HKLL operators will drop out after the trace is taken due to the isometry of HKLL.}
    \label{fig:SO_line1}
\end{figure}

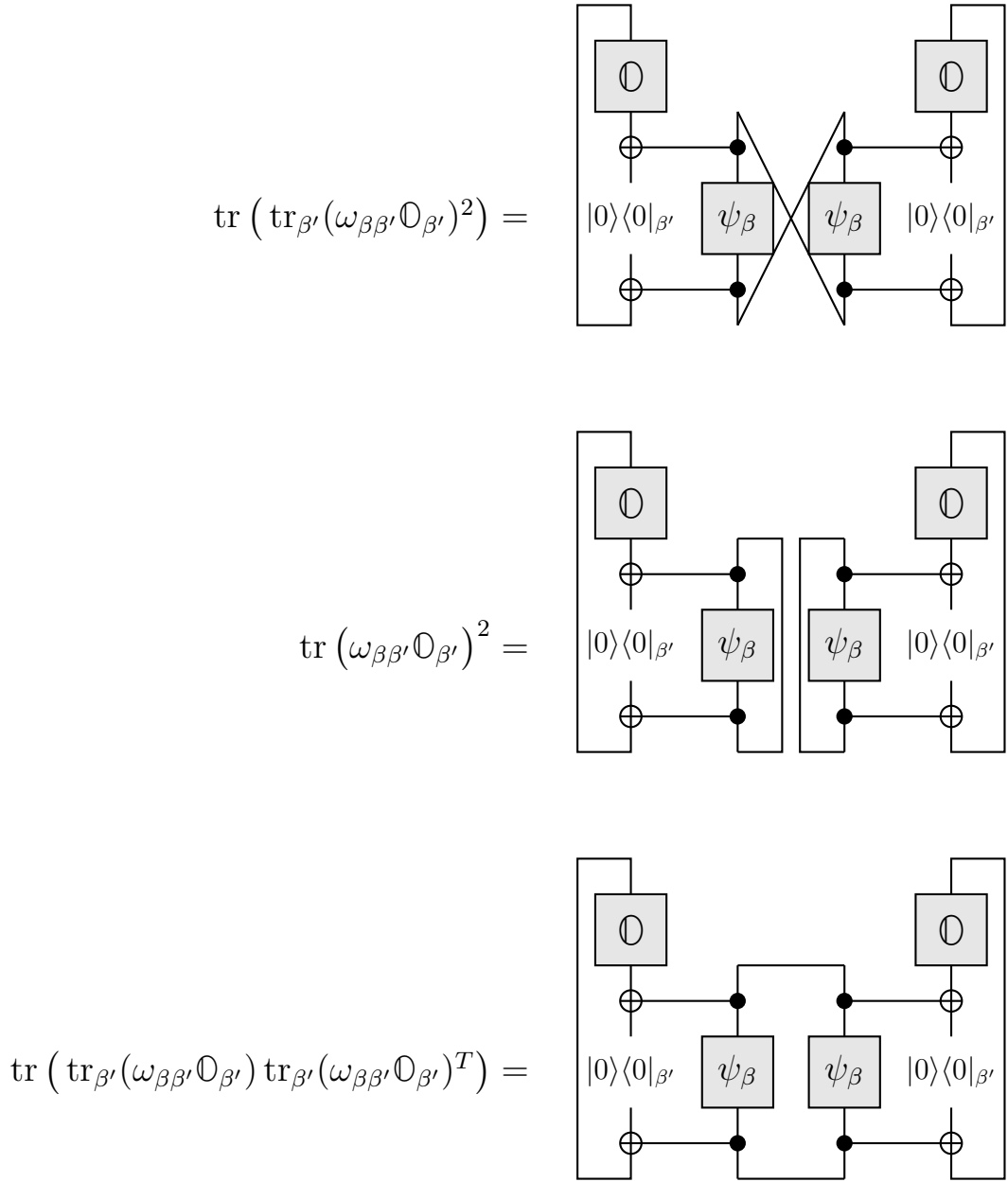
\begin{figure}
    \centering
    \begin{tikzpicture}[thick, scale=1, every node/.style={transform shape}]

\newcommand{\CNOT}[4]{
    \draw[fill=black] (#1,#2) circle (0.1);
    \draw (#1,#2) -- (#3,#4);
    \draw[fill=white] (#3,#4) circle (0.15);
    \draw (#3+0.15,#4) -- (#3-0.15,#4);
    \draw (#3,#4+0.15) -- (#3,#4-0.15);
}

\begin{scope}[shift={(0,0)}]
    \draw[fill=gray!20] (0.25,0.5) rectangle (1.25,-0.5);
    \node[scale=1.2] at (0.75,0) {$\psi_\beta$};
    \draw[fill=gray!20] (-0.25,0.5) rectangle (-1.25,-0.5);
    \node[scale=1.2] at (-0.75,0) {$\psi_\beta$};
    \draw (0.75,0.5) -- (0.75,1.5);
    \draw (0.75,-0.5) -- (0.75,-1.5);
    \draw (-0.75,0.5) -- (-0.75,1.5);
    \draw (-0.75,-0.5) -- (-0.75,-1.5);
    \node[scale=1] at (2.25,0) {$|0\rangle\langle0|_{\beta'}$};
    \node[scale=1] at (-2.25,0) {$|0\rangle\langle0|_{\beta'}$};
    \draw (2.25,0.5) -- (2.25,1.5);
    \draw (2.25,2.5) -- (2.25,3) -- (3,3) -- (3,-1.5) -- (2.25,-1.5) -- (2.25,-0.5);
    \draw (-2.25,0.5) -- (-2.25,1.5);
    \draw (-2.25,2.5) -- (-2.25,3) -- (-3,3) -- (-3,-1.5) -- (-2.25,-1.5) -- (-2.25,-0.5);
    \draw[fill=gray!20] (1.75,1.5) rectangle (2.75,2.5);
    \node[scale=1.4] at (2.25,2) {$\mathbb{O}$};
    \draw[fill=gray!20] (-1.75,1.5) rectangle (-2.75,2.5);
    \node[scale=1.4] at (-2.25,2) {$\mathbb{O}$};
    \CNOT{0.75}{1}{2.25}{1};
    \CNOT{0.75}{-1}{2.25}{-1};
    \CNOT{-0.75}{1}{-2.25}{1};
    \CNOT{-0.75}{-1}{-2.25}{-1};
    \draw (0.75,1.5) -- (-0.75,-1.5);
    \draw (-0.75,1.5) -- (0.75,-1.5);
    \node[scale=1.2,anchor=east] at (-3.5,0) {$\tr \big( \tr_{\beta'} ( \omega_{\beta\beta'} \mathbb{O}_{\beta'} )^2 \big) = $};
\end{scope}

\begin{scope}[shift={(0,-6)}]
    \draw[fill=gray!20] (0.25,0.5) rectangle (1.25,-0.5);
    \node[scale=1.2] at (0.75,0) {$\psi_\beta$};
    \draw[fill=gray!20] (-0.25,0.5) rectangle (-1.25,-0.5);
    \node[scale=1.2] at (-0.75,0) {$\psi_\beta$};
    \draw (0.75,0.5) -- (0.75,1.5);
    \draw (0.75,-0.5) -- (0.75,-1.5);
    \draw (-0.75,0.5) -- (-0.75,1.5);
    \draw (-0.75,-0.5) -- (-0.75,-1.5);
    \node[scale=1] at (2.25,0) {$|0\rangle\langle0|_{\beta'}$};
    \node[scale=1] at (-2.25,0) {$|0\rangle\langle0|_{\beta'}$};
    \draw (2.25,0.5) -- (2.25,1.5);
    \draw (2.25,2.5) -- (2.25,3) -- (3,3) -- (3,-1.5) -- (2.25,-1.5) -- (2.25,-0.5);
    \draw (-2.25,0.5) -- (-2.25,1.5);
    \draw (-2.25,2.5) -- (-2.25,3) -- (-3,3) -- (-3,-1.5) -- (-2.25,-1.5) -- (-2.25,-0.5);
    \draw[fill=gray!20] (1.75,1.5) rectangle (2.75,2.5);
    \node[scale=1.4] at (2.25,2) {$\mathbb{O}$};
    \draw[fill=gray!20] (-1.75,1.5) rectangle (-2.75,2.5);
    \node[scale=1.4] at (-2.25,2) {$\mathbb{O}$};
    \CNOT{0.75}{1}{2.25}{1};
    \CNOT{0.75}{-1}{2.25}{-1};
    \CNOT{-0.75}{1}{-2.25}{1};
    \CNOT{-0.75}{-1}{-2.25}{-1};
    \draw (0.75,1.5) -- (0.125,1.5) -- (0.125,-1.5) -- (0.75,-1.5);
    \draw (-0.75,1.5) -- (-0.125,1.5) -- (-0.125,-1.5) -- (-0.75,-1.5);
    \node[scale=1.2,anchor=east] at (-3.5,0) {$\tr \big( \omega_{\beta\beta'} \mathbb{O}_{\beta'} \big)^2 = $};
\end{scope}

\begin{scope}[shift={(0,-12)}]
    \draw[fill=gray!20] (0.25,0.5) rectangle (1.25,-0.5);
    \node[scale=1.2] at (0.75,0) {$\psi_\beta$};
    \draw[fill=gray!20] (-0.25,0.5) rectangle (-1.25,-0.5);
    \node[scale=1.2] at (-0.75,0) {$\psi_\beta$};
    \draw (0.75,0.5) -- (0.75,1.5);
    \draw (0.75,-0.5) -- (0.75,-1.5);
    \draw (-0.75,0.5) -- (-0.75,1.5);
    \draw (-0.75,-0.5) -- (-0.75,-1.5);
    \node[scale=1] at (2.25,0) {$|0\rangle\langle0|_{\beta'}$};
    \node[scale=1] at (-2.25,0) {$|0\rangle\langle0|_{\beta'}$};
    \draw (2.25,0.5) -- (2.25,1.5);
    \draw (2.25,2.5) -- (2.25,3) -- (3,3) -- (3,-1.5) -- (2.25,-1.5) -- (2.25,-0.5);
    \draw (-2.25,0.5) -- (-2.25,1.5);
    \draw (-2.25,2.5) -- (-2.25,3) -- (-3,3) -- (-3,-1.5) -- (-2.25,-1.5) -- (-2.25,-0.5);
    \draw[fill=gray!20] (1.75,1.5) rectangle (2.75,2.5);
    \node[scale=1.4] at (2.25,2) {$\mathbb{O}$};
    \draw[fill=gray!20] (-1.75,1.5) rectangle (-2.75,2.5);
    \node[scale=1.4] at (-2.25,2) {$\mathbb{O}$};
    \CNOT{0.75}{1}{2.25}{1};
    \CNOT{0.75}{-1}{2.25}{-1};
    \CNOT{-0.75}{1}{-2.25}{1};
    \CNOT{-0.75}{-1}{-2.25}{-1};
    \draw (0.75,1.5) -- (-0.75,1.5);
    \draw (-0.75,-1.5) -- (0.75,-1.5);
    \node[scale=1.2,anchor=east] at (-3.5,0) {$\tr \big( \tr_{\beta'} ( \omega_{\beta\beta'} \mathbb{O}_{\beta'} ) \tr_{\beta'} (\omega_{\beta\beta'} \mathbb{O}_{\beta'} )^T \big) = $};
\end{scope}
    
\end{tikzpicture}
    \caption{Circuit diagram depictions of the $\mathbb{O}_{\beta'}$ dependent terms in the second equality of (\ref{eq:expRbaby}) after the average over $O$ has been taken.}
    \label{fig:SO_line2}
\end{figure}

Let us now consider other choices of $\tilde{\swap}_\beta$ to determine if there is an alternative that $\beta$ can use to more accurately predict the SWAP test. We will continue to assume that $\tilde{\swap}_\beta$ acts as $\tilde{\swap}_\text{AdS}$ on the $\hs_A \subset \hs_{A\beta'}$ subsystem, but we allow for more general operations $\mathbb{O}$ on $\beta'$:
\begin{equation} \label{eq:S_beta_O}
    \tilde{\swap}_\beta^{(\mathbb{O})} \equiv \tilde{\swap}_\text{AdS} \otimes \mathbb{O}_{\beta'}.
\end{equation}
The expectation of this more general operator is given by
\begin{align*} 
    \int dO\, \langle \tilde{\swap}_\beta^{(\mathbb{O})} \rangle_{V_\beta|\psi_1\rangle} &= \int dO\, \tr \big[ V_\beta (\psi_1 \otimes \psi_\alpha \otimes \psi_\beta)^{\otimes2} V^\dagger_\beta \big( \tilde{\swap}_\text{AdS} \otimes \mathbb{O}_{\beta'} \big) \big] \\
        &= \frac{d_b}{d_b + 2} \Big[ \tr \big( \tr_{\beta'} ( \omega_{\beta\beta'} \mathbb{O}_{\beta'} )^2 \big) + \tr \big( \omega_{\beta\beta'} \mathbb{O}_{\beta'} \big)^2 \tr \big( \psi_{1,M_a}^2 \big) \\
        &\hspace{2cm} + \tr \big( \tr_{\beta'} ( \omega_{\beta\beta'} \mathbb{O}_{\beta'} ) \tr_{\beta'} (\omega_{\beta\beta'} \mathbb{O}_{\beta'} )^T \big) \tr \big( \psi_1 \psi_1^{T_{M_b}} \big) \Big] \numberthis \label{eq:expRbaby}
\end{align*}
To assist in interpreting this result, figure \ref{fig:SO_line1} provides a circuit diagram for the contents of the trace on the right hand side of the first line, and figure \ref{fig:SO_line2} provides diagrams for the three $\mathbb{O}_{\beta'}$ dependent terms in the second equality. Note that these $\mathbb{O}_{\beta'}$ terms are the only difference between (\ref{eq:expSbaby}) and (\ref{eq:expRbaby}) -- the $\psi_1$ dependence is the same in both.

EGH's choice of $\mathbb{O}_{\beta'} = \id_{\beta'}$ sets the $\beta$ dependence of the second term in (\ref{eq:expRbaby}) to 1, reproducing (\ref{eq:EGH8}). Note that this choice leads to exact agreement between (\ref{eq:EGH8}) and (\ref{eq:expSbaby}) at subleading order in $\psi_1$. If we instead choose $\mathbb{O}_{\beta'}$ so that the leading terms in $\psi_1$ agree, we could reduce the error $\Delta$. Indeed, swapping the clone $\beta'$ by setting $\mathbb{O}_{\beta'} = \tilde\swap_{\beta'}$, we find the first term in (\ref{eq:expRbaby}) reduces to
\begin{equation}
    \tr \big( \tr_{\beta'} ( \omega_{\beta\beta'} \tilde{\swap}_{\beta'} )^2 \big) = 1,
\end{equation}
which matches the leading order term in (\ref{eq:expSbaby}). The full expectation value for this choice of reconstruction is
\begin{equation}
    \int dO\, \langle \tilde{\swap}_\text{AdS} \otimes \tilde{\swap}_{\beta'} \rangle_{V_\beta |\psi_1\rangle} = \frac{d_b}{d_b + 2} \Big[ 1 + \tr \big( \omega_{\beta'}^2 \big) \tr \big( \psi_{1,M_a}^2 \big) + \tr \big( \omega \omega^{T_\beta} \big) \tr \big( \psi_1 \psi_1^{T_{M_b}} \big) \Big].
\end{equation}
This differs from (\ref{eq:expSbaby}) by
\begin{align*} 
    \overline{\Delta}_\swap &= \frac{d_b}{d_b + 2} \Big[ \Big( 1 - \tr \big( \omega_{\beta'}^2 \big) \Big) \tr \big( \psi_{1,M_a}^2 \big) + \Big( \tr \big( \psi_\beta \psi_\beta^T \big) - \tr \big( \omega \omega^{T_\beta} \big) \Big) \tr \big( \psi_1 \psi_1^{T_{M_b}} \big) \Big] \\
        &\approx \Big( 1 - \tr \big( \omega_{\beta'}^2 \big) \Big) \tr \big( \psi_{1,M_a}^2 \big) \numberthis \label{eq:deltaS},
\end{align*}
where again we have used a bar to emphasize that an average over $O$ has been taken. In the final line, we have again taken the large $d_b$ limit and dropped terms subleading in $\psi_1$. Notice that this error is smaller than $\overline{\Delta}_\id$ in (\ref{eq:delta1}) by a factor of $\tr\big(\psi_{1,M_a}^2\big) = e^{-S_2(\psi_{1,M_a})}$. By making the entanglement between $M_a$ and $M_b$ in $|\psi_1\rangle$ larger, we can decrease the average error between the predictions of $\tilde\swap_\text{AdS} \otimes \tilde\swap_{\beta'}$ and $\swap_\text{baby}$. 

\paragraph{Exact reconstruction in a toy qubit model.} So far we have only demonstrated that $\overline{\Delta}_\swap < \overline{\Delta}_\id$. However, this does not imply $\overline{\Delta^2_\swap} < \overline{\Delta^2_\id}$ -- it could be that $\tilde\swap_\text{AdS} \otimes \tilde\swap_{\beta'}$ has a larger variance, making it a worse prediction for individual choices of $O$.\footnote{We thank Chris Waddell for discussions on this point.} Instead of computing the variance directly, let us compute $\Delta_\swap$ and $\Delta_\id$ exactly for a particular choice of $O$. While this does not constitute a proof of the improved prediction by $\tilde\swap_\text{AdS} \otimes \tilde\swap_{\beta'}$, we take it to give strong evidence.

We again consider a qubit model where
\begin{equation}
    |\psi_1\rangle_{M_a M_b} = |\Phi^+\rangle_{M_a M_b}, \qquad \prescript{}{b}{\langle0|}O = \prescript{}{\beta M_b}{\langle\Phi^+|}.
\end{equation}
Recall that the AR condition is satisfied in this model for any $|\psi_2\rangle_{M_a} = |\psi_\beta\rangle$. Therefore, we expect $\swap_\text{baby}$ to have an expectation value of 1 on $|\psi_1\rangle$:
\begin{equation} \label{eq:model_Sbaby}
    \langle \swap_\text{baby} \rangle_{|\Phi^+\rangle} = 16 \tr \Big[ \big( \Phi^+_{M_aM_b} \otimes \psi_\alpha \otimes \psi_\beta \big)^{\otimes 2} \big( \swap_\text{AdS} \otimes |\Phi^+\rangle\langle\Phi^+|_{\beta M_b} \big) \Big] = \tr\big(\psi_\beta^2\big) = 1,
\end{equation}
where $\Phi^+ \equiv |\Phi^+\rangle\langle\Phi^+|$. This is indeed 1 since the $\beta$ observer is taken to be in the pure state $|\psi_\beta\rangle$. For $\tilde{\swap}_\beta$ to be a good reconstruction of $\swap_\text{baby}$, we expect its expectation value on $V_\beta |\Phi^+\rangle$ to closely match this.

First consider a candidate reconstruction given by EGH's operator $\tilde{\swap}_\text{AdS}\otimes\id_{\beta'}$. In this toy model, the expectation value is
\begin{align*}
    \langle \tilde{\swap}_\text{AdS}\otimes\id_{\beta'} \rangle_{V_\beta |\Phi^+\rangle} &= \tr \Big[ V_\beta \big(\Phi^+_{M_a M_b} \otimes \psi_\alpha \otimes \psi_\beta \big)^{\otimes 2} V^\dagger_\beta \big(\tilde{\swap}_\text{AdS} \otimes \mathbb{1}_{\beta'} \big) \Big] \\
        &= \tr \Big[ (V\otimes\id_{\beta'}) \big(\Phi^+_{M_a M_b} \otimes \psi_\alpha \otimes \omega_{\beta\beta'} \big)^{\otimes 2} (V\otimes\id_{\beta'})^\dagger \big(\tilde{\swap}_\text{AdS} \otimes \mathbb{1}_{\beta'} \big) \Big] \\
        &= \tr \Big[ \big(\Phi^+_{M_a M_b} \otimes \psi_\alpha \otimes \omega_{\beta\beta'} \big)^{\otimes 2} (V\otimes\id_{\beta'})^\dagger \big(\tilde{\swap}_\text{AdS} \otimes \mathbb{1}_{\beta'} \big) (V\otimes\id_{\beta'}) \Big] \\
        &= 16 \tr \Big[ \big(\Phi^+_{M_a M_b} \otimes \psi_\alpha \otimes \omega_{\beta\beta'} \big)^{\otimes 2} \big( \swap_\text{AdS} \otimes |\Phi^+\rangle\langle\Phi^+|_{\beta M_b} \otimes \id_{\beta'} \big) \Big] \\
        &= \tr\big( \omega_\beta^2 \big), \numberthis \label{eq:expId_model}
\end{align*}
where we applied the cloning operation to $\psi_\beta$ in the second equality such that
\begin{equation}
    V_\beta |\psi_\beta\rangle = (V\otimes\id_{\beta'}) |\omega\rangle_{\beta\beta'}.
\end{equation}
We also used the cyclicity of the trace in the third equality and equation (\ref{eq:S_baby}) in the fourth. Finally, we find that the expectation value is just the purity of $\beta$ after the clone is traced out. This is generically less than 1, with error
\begin{equation}
    \Delta_\id = \langle \swap_\text{baby} \rangle_{|\Phi^+\rangle} - \langle \tilde{\swap}_\text{AdS}\otimes\id_{\beta'} \rangle_{V_\beta |\Phi^+\rangle} = 1 - \tr\big(\omega_\beta^2\big)
\end{equation}
exactly matching the approximate error found in (\ref{eq:delta1}) above.

Let us now consider a reconstruction given by $\tilde{\swap}_\text{AdS}\otimes\tilde{\swap}_{\beta'}$, which we expect from (\ref{eq:deltaS}) to give a better approximation of $\langle\swap_\text{baby}\rangle$. Following the same steps as in (\ref{eq:expId_model}), we find the expectation value of this new operator to be
\begin{align*}
    \langle \tilde{\swap}_\text{AdS}\otimes\tilde{\swap}_{\beta'} \rangle_{V_\beta |\Phi^+\rangle} &= \tr \Big[ V_\beta \big(\Phi^+_{M_a M_b} \otimes \psi_\alpha \otimes \psi_\beta \big)^{\otimes 2} V^\dagger_\beta \big(\tilde{\swap}_\text{AdS} \otimes \tilde{\swap}_{\beta'} \big) \Big] \\
        &= \dots \\
        &= 16 \tr \Big[ \big(\Phi^+_{M_a M_b} \otimes \psi_\alpha \otimes \omega_{\beta\beta'} \big)^{\otimes 2} \big( \swap_\text{AdS} \otimes |\Phi^+\rangle\langle\Phi^+|_{\beta M_b} \otimes \tilde\swap_{\beta'} \big) \Big] \\
        &= \tr\big( \omega_{\beta\beta'}^2 \big) = 1. \numberthis
\end{align*}
Since $|\psi_\beta\rangle$ is in a pure state, the cloned state $|\omega\rangle_{\beta\beta'}$ is also pure. Therefore the expectation value of $\tilde{\swap}_\text{AdS}\otimes\tilde{\swap}_{\beta'}$ is 1, exactly matching (\ref{eq:model_Sbaby}) with zero error!

In fact, we find that the induced operator map $V^*_\beta$ takes $\tilde{\swap}_\text{AdS}\otimes\tilde{\swap}_{\beta'}$ back to $\swap_\text{baby}$, satisfying (\ref{eq:exactSbeta}) for an exact reconstruction. Using $\Phi^+_{M_a M_b}$ as the projector on to the $|\psi_1\rangle_{M_a M_b} = |\Phi^+\rangle_{M_a M_b}$ subspace we have been considering in this toy model, we find
\begin{equation}
    \Phi^+_{M_a M_b} V^\dagger_\beta (\tilde{\swap}_\text{AdS} \otimes \tilde\swap_{\beta'}) V_\beta \Phi^+_{M_a M_b} =  \Phi^+_{M_a M_b} \otimes \swap_\beta \otimes \id_\alpha = \Phi^+_{M_a M_b} \swap_\text{baby} \Phi^+_{M_a M_b}
\end{equation}
where $\swap_\beta$ is a SWAP on the state of the $\beta$ observer in the effective description. In this toy model, $\tilde\swap_\beta = \tilde{\swap}_\text{AdS}\otimes\tilde{\swap}_{\beta'}$ is an exact reconstruction of $\swap_\text{baby}$, and there is no error in $\beta$'s prediction for the SWAP test.

\paragraph{How does $\beta$ describe the bulk?} Here there is only one state on which the $\beta$ observer can measure the SWAP test, namely $V_\beta |\psi_1\rangle$. There is no holographic map that takes $|\psi_2\rangle \in \hs_a$ to the $\beta$ observer's Hilbert space $\hs_{A\beta'}$. Therefore, the $\beta$ observer \textit{must} describe their world with the closed universe in which they live. 

Like EGH, we find that the $\beta$ observer's description is generically limited due to the non-isometry of $V_\beta$. However, we find that $\tilde\swap_\beta = \tilde{\swap}_\text{AdS}\otimes\tilde{\swap}_{\beta'}$ provides a better approximation of the SWAP test for the $\beta$ observer, with an error improved by additional entanglement between the matter in the AdS bulk and baby universe. This is reminiscent of the results from \cite{antonini_cosmology_2023,antonini_baby_2025}, which found that tensor networks modeling a holographic map from the baby universe to the AdS boundaries preserved more of the information contained in the baby universe when the entanglement with the AdS bulk matter was increased. Furthermore, we find that this new operator provides an exact reconstruction in a toy model, demonstrating that the error is not fundamental.

\section{CO rules for observers} \label{sec:CO}

In this section we consider how the predictions of $\alpha$ and $\beta$ change if the CO rules \cite{akers_observers_2025} are used. We begin with a brief review of these rules before using them to construct new definitions of $\hat{V}_\alpha$, $V_\alpha$, and $V_\beta$.

Motivated by the assumption that observers are already a part of the fundamental description, the CO rules remove any part of the holographic map that acts on the observer. We emphasize that this does not alter the bulk state, only the holographic map to the fundamental description. This modification requires the holographic map to have some notion of locality in order to distinguish the part of the holographic map acting on the observer from the remainder acting on their environment. 

In \cite{akers_observers_2025}, tensor networks were used to model this locality, in which case the CO rules instruct us to remove any tensors acting on the observer. The resulting observer-modified holographic map then acts as the identity on the observer subsystem. Furthermore, the excision of these tensors leaves dangling in-plane legs across the boundary between the observer and the remainder of the network. The resulting fundamental Hilbert space is enlarged by these additional boundary degrees of freedom, which we label here as $\partial$. Following the PEPS formalism \cite{verstraete_renormalization_2004},\footnote{For reviews of the PEPS formalism, see for example \cite{hayden_holographic_2016,akers_background_2024}.} these dangling legs are taken to be maximally entangled with inputs $\partial'$ to neighboring tensors, representing the spatial connectivity of the bulk across the boundary. 

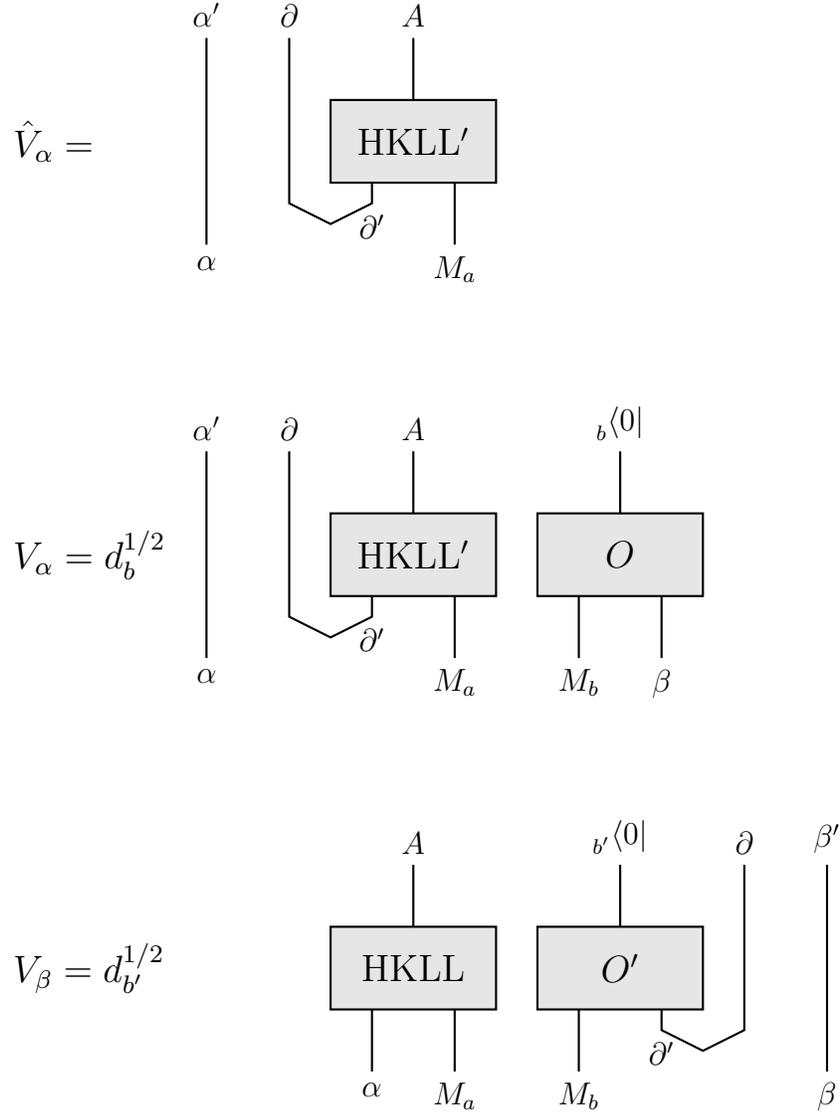
\begin{figure}
    \centering
    \begin{tikzpicture}[thick,scale=1.1]

\begin{scope}[xscale=-1]
    \draw (0,0.25) -- (0,1);
    \node[anchor=north] at (0,0.25) {$M_a$};
    \draw (1,1) -- (1,0.75) -- (1.5,0.5) -- (2,0.75) -- (2,2.75);
    \node[anchor=north] at (1,0.75) {$\partial'$};
    \node[anchor=south] at (2,2.75) {$\partial$};
    \draw (3,0.25) -- (3,2.75);
    \node[anchor=north] at (3,0.25) {$\alpha$};
    \node[anchor=south] at (3,2.75) {$\alpha'$};
    \draw[fill=gray!20] (-0.5,1) rectangle (1.5,2);
    \node[scale=1.2] at (0.5,1.5) {HKLL$'$};
    \draw (0.5,2) -- (0.5,2.75);
    \node[anchor=south] at (0.5,2.75) {$A$};
    \node[anchor=west,scale=1.2] at (5.5, 1.5) {$\hat{V}_\alpha =$};
\end{scope}

\begin{scope}[shift={(0,-5)},xscale=-1]
    \draw (-2.5,0.25) -- (-2.5,1);
    \node[anchor=north] at (-2.5,0.25) {$\beta$};
    \draw (-1.5,0.25) -- (-1.5,1);
    \node[anchor=north] at (-1.5,0.25) {$M_b$};
    \draw[fill=gray!20] (-3,1) rectangle (-1,2);
    \node[scale=1.2] at (-2,1.5) {$O$};
    \draw (-2,2) -- (-2,2.75);
    \node[anchor=south] at (-2,2.75) {$\prescript{}{b}{\langle0|}$};
    \draw (0,0.25) -- (0,1);
    \node[anchor=north] at (0,0.25) {$M_a$};
    \draw (1,1) -- (1,0.75) -- (1.5,0.5) -- (2,0.75) -- (2,2.75);
    \node[anchor=north] at (1,0.75) {$\partial'$};
    \node[anchor=south] at (2,2.75) {$\partial$};
    \draw (3,0.25) -- (3,2.75);
    \node[anchor=north] at (3,0.25) {$\alpha$};
    \node[anchor=south] at (3,2.75) {$\alpha'$};
    \draw[fill=gray!20] (-0.5,1) rectangle (1.5,2);
    \node[scale=1.2] at (0.5,1.5) {HKLL$'$};
    \draw (0.5,2) -- (0.5,2.75);
    \node[anchor=south] at (0.5,2.75) {$A$};
    \node[anchor=west,scale=1.2] at (5.5, 1.5) {$V_\alpha = d_b^{1/2}$};
\end{scope}

\begin{scope}[shift={(0,-10)},xscale=-1]
    \draw (-4.5,0.25) -- (-4.5,2.75);
    \node[anchor=north] at (-4.5,0.25) {$\beta$};
    \node[anchor=south] at (-4.5,2.75) {$\beta'$};
    \draw (-2.5,1) -- (-2.5,0.75) -- (-3,0.5) -- (-3.5,0.75) -- (-3.5,2.75);
    \node[anchor=north] at (-2.5,0.75) {$\partial'$};
    \node[anchor=south] at (-3.5,2.75) {$\partial$};
    \draw (-1.5,0.25) -- (-1.5,1);
    \node[anchor=north] at (-1.5,0.25) {$M_b$};
    \draw[fill=gray!20] (-3,1) rectangle (-1,2);
    \node[scale=1.2] at (-2,1.5) {$O'$};
    \draw (-2,2) -- (-2,2.75);
    \node[anchor=south] at (-2,2.75) {$\prescript{}{b'}{\langle0|}$};
    \draw (0,0.25) -- (0,1);
    \node[anchor=north] at (0,0.25) {$M_a$};
    \draw (1,0.25) -- (1,1);
    \node[anchor=north] at (1,0.25) {$\alpha$};
    \draw[fill=gray!20] (-0.5,1) rectangle (1.5,2);
    \node[scale=1.2] at (0.5,1.5) {HKLL};
    \draw (0.5,2) -- (0.5,2.75);
    \node[anchor=south] at (0.5,2.75) {$A$};
    \node[anchor=west,scale=1.2] at (5.5, 1.5) {$V_\beta = d_{b'}^{1/2}$};
\end{scope}
    
\end{tikzpicture}
    \caption{Circuit diagrams for the three observer-modified holographic maps constructed using the CO rules. Primes on HKLL$'$ and $O'$ denote modified operators created by excising the parts of HKLL or $O$ that act on the observer, leaving $\id$ acting on the observer and additional dangling legs $\partial$ on the boundary between the observer and their environment.}
    \label{fig:CO_rules}
\end{figure}

Of course, $V_\text{HKLL}$ and $V$ shown in figure \ref{fig:map2A} lack the locality needed to directly apply these rules. Instead, we will imagine both are modeled by tensor networks and construct models of $\hat{V}_\alpha$, $V_\alpha$, and $V_\beta$. For example, we will model the removal of tensors acting on $\alpha$ by replacing the operator $\text{HKLL}:M_a\alpha \to A$ in figure \ref{fig:map2A} with a modified operator $\text{HKLL}':M_a\partial' \to A$. This leaves $\id$ acting on $\alpha$ (relabeling it to $\alpha'$) and dangling legs $\partial$ maximally entangled with the inputs $\partial'$ to HKLL$'$. Depictions of $\hat{V}_\alpha$ and $V_\alpha$ constructed in this way are shown in figure \ref{fig:CO_rules}. We will assume that $\text{HKLL}'$ remains isometric, as removing tensors from the network modeling HKLL should not introduce any non-isometries. As a result, $\hat{V}_\alpha$ is again isometric.

Similarly, we apply the CO rules to the $\beta$ observer by replacing $O$ acting on $M_b\beta$ with $O'$ acting on $M_b\partial'$. This again leaves $\id$ acting on $\beta$ (relabeling it to $\beta'$) and dangling legs $\partial$ maximally entangled with the inputs $\partial'$. Figure \ref{fig:CO_rules} also includes a circuit diagram representation of $V_\beta$ constructed in this way. Again, we will assume that $O'$ retains the orthogonal property of $O$.

\subsection{\texorpdfstring{$\alpha$}{alpha} observer} \label{sec:CO_alpha}
We now use $\hat{V}_\alpha$ constructed using the CO rules to define $\tilde\swap_\alpha$ according to (\ref{eq:S_alpha}):
\begin{equation} \label{eq:S_alpha_CO}
    \tilde{\swap}_\alpha = 
    \vcenter{\hbox{\begin{tikzpicture}[thick,scale=0.8]

\draw (0,1) -- (0,0.25) -- (-1.5,-0.5) -- (-1.5,-1.25);
\draw (-1.5,1) -- (-1.5,0.25) -- (0,-0.5) -- (0,-1.25);
\node[scale=0.8,anchor=west] at (-1.5,-0.875) {$M_a$};
\draw (1,1) -- (1,0.75) -- (1.5,0.5) -- (2,0.75) -- (2,2.75);
\node[anchor=south] at (2,2.75) {$\partial$};
\draw (-2.5,1) -- (-2.5,0.75) -- (-3,0.5) -- (-3.5,0.75) -- (-3.5,2.75);
\node[anchor=south] at (-3.5,2.75) {$\partial$};
\draw (1,-1.25) -- (1,-1) -- (1.5,-0.75) -- (2,-1) -- (2,-3);
\node[anchor=south] at (1,-1) {$\partial'$};
\node[anchor=north] at (2,-3) {$\partial$};
\draw (-2.5,-1.25) -- (-2.5,-1) -- (-3,-0.75) -- (-3.5,-1) -- (-3.5,-3);
\node[anchor=south] at (-2.5,-1) {$\partial'$};
\node[anchor=north] at (-3.5,-3) {$\partial$};
\draw (3,-3) -- (3,2.75);
\node[anchor=south] at (3,2.75) {$\alpha'$};
\node[anchor=north] at (3,-3) {$\alpha'$};
\draw (-4.5,-3) -- (-4.5,2.75);
\node[anchor=south] at (-4.5,2.75) {$\alpha'$};
\node[anchor=north] at (-4.5,-3) {$\alpha'$};
\draw[fill=gray!20] (-0.5,1) rectangle (1.5,2);
\node at (0.5,1.5) {HKLL$'$};
\draw[fill=gray!20] (-3,1) rectangle (-1,2);
\node at (-2,1.5) {HKLL$'$};
\draw[fill=gray!20] (-0.5,-2.25) rectangle (1.5,-1.25);
\node at (0.5,-1.75) {HKLL$'^\dagger$};
\draw[fill=gray!20] (-3,-2.25) rectangle (-1,-1.25);
\node at (-2,-1.75) {HKLL$'^\dagger$};
\draw (0.5,2) -- (0.5,2.75);
\node[anchor=south] at (0.5,2.75) {$A$};
\draw (-2,2) -- (-2,2.75);
\node[anchor=south] at (-2,2.75) {$A$};
\draw (0.5,-2.25) -- (0.5,-3);
\node[anchor=north] at (0.5,-3) {$A$};
\draw (-2,-2.25) -- (-2,-3);
\node[anchor=north] at (-2,-3) {$A$};

\draw[decorate,decoration={brace,amplitude=6pt,mirror}] (4,-0.5) -- (4,0.5);
\node[scale=1,anchor=west] at (4.25,0) {$\swap_\text{AdS}$};

\draw[decorate,decoration={brace,amplitude=6pt,mirror}] (4,0.6) -- (4,2.75);
\node[scale=1,anchor=west] at (4.25,1.675) {$\hat{V}_\alpha$};

\draw[decorate,decoration={brace,amplitude=6pt}] (4,-0.6) -- (4,-3);
\node[scale=1,anchor=west] at (4.25,-1.675) {$\hat{V}^\dagger_\alpha$};
        
\end{tikzpicture}}}
\end{equation}
When we used the HUZ rules in section \ref{sec:HUZ_alpha}, we found that $\tilde\swap_\alpha$ reduces to EGH's choice of $\tilde\swap_\text{AdS}\otimes\id_{\alpha'}$. We see that this is not the case with the CO rules -- while (\ref{eq:S_alpha_CO}) does act as $\id$ on $\alpha'$, the modification of HKLL to HKLL$'$ means that the remainder does not reduce to $\tilde\swap_\text{AdS}$ acting on $A$. Therefore, $\tilde\swap_\text{AdS} \otimes \id_{\alpha'}$ is not the correct way to measure the SWAP test according to $\alpha$ when the CO rules are used. Instead, equation (\ref{eq:exp_psi2}) and the isometry of $\hat{V}_\alpha$ guarantee that the expectation value of $\tilde\swap_\alpha$ defined in (\ref{eq:S_alpha_CO}) will agree with the SWAP test $\swap_\text{AdS}$ on the bulk state $|\psi_2\rangle$. Using $\tilde\swap_\alpha$, we find that the $\alpha$ observer incorporated via the CO rules can describe the bulk without a baby universe.

What about $|\psi_1\rangle$ for the bulk with a baby universe? Recall that the validity of (\ref{eq:exp_psi1}) -- which tells us that $\tilde\swap_\alpha$ defined in (\ref{eq:S_alpha_CO}) will also match $\swap_\text{baby}$ on $|\psi_1\rangle$ -- requires $V_\alpha^\dagger \hat{V}_\alpha = V^\dagger V_\text{HKLL}$. We verify this holds for the CO rules using $\hat{V}_\alpha$ and $V_\alpha$ shown in figure \ref{fig:CO_rules}:
\begin{equation}
    V_\alpha^\dagger \hat{V}_\alpha = 
    \vcenter{\hbox{\begin{tikzpicture}[thick,scale=0.8]

\draw (0,0.25) -- (0,1);
\node[anchor=north] at (0,0.25) {$M_a$};
\draw (1,1) -- (1,0.75) -- (1.5,0.5) -- (2,0.75) -- (2,2.75);
\node[anchor=north] at (1,0.75) {$\partial'$};
\draw (3,0.25) -- (3,2.75);
\node[anchor=north] at (3,0.25) {$\alpha$};
\draw[fill=gray!20] (-0.5,1) rectangle (1.5,2);
\node at (0.5,1.5) {HKLL$'$};
\draw (0.5,2) -- (0.5,2.75);

\begin{scope}[shift={(0,4.75)},yscale=-1]
    \draw (-2.5,0.25) -- (-2.5,1);
    \node[anchor=south] at (-2.5,0.25) {$\beta$};
    \draw (-1.5,0.25) -- (-1.5,1);
    \node[anchor=south] at (-1.5,0.25) {$M_b$};
    \draw[fill=gray!20] (-3,1) rectangle (-1,2);
    \node at (-2,1.5) {$O^T$};
    \draw (-2,2) -- (-2,2.75);
    \node[anchor=north] at (-2,2.75) {$|0\rangle_b$};
    \draw (0,0.25) -- (0,1);
    \node[anchor=south] at (0,0.25) {$M_a$};
    \draw (1,1) -- (1,0.75) -- (1.5,0.5) -- (2,0.75) -- (2,2);
    \node[anchor=south] at (1,0.75) {$\partial'$};
    \draw (3,0.25) -- (3,2);
    \node[anchor=south] at (3,0.25) {$\alpha$};
    \draw[fill=gray!20] (-0.5,1) rectangle (1.5,2);
    \node at (0.5,1.5) {HKLL$'^\dagger$};
\end{scope}

\end{tikzpicture}}}
    = O^T|0\rangle_b \otimes \id_{M_a\alpha} = 
    \vcenter{\hbox{\begin{tikzpicture}[thick,scale=0.8]

\draw (0,0.25) -- (0,1);
\node[anchor=north] at (0,0.25) {$M_a$};
\draw (1,0.25) -- (1,1);
\node[anchor=north] at (1,0.25) {$\alpha$};
\draw[fill=gray!20] (-0.5,1) rectangle (1.5,2);
\node at (0.5,1.5) {HKLL};
\draw (0.5,2) -- (0.5,2.75);

\begin{scope}[shift={(0,4.75)},yscale=-1]
    \draw (-2.5,0.25) -- (-2.5,1);
    \node[anchor=south] at (-2.5,0.25) {$\beta$};
    \draw (-1.5,0.25) -- (-1.5,1);
    \node[anchor=south] at (-1.5,0.25) {$M_b$};
    \draw[fill=gray!20] (-3,1) rectangle (-1,2);
    \node at (-2,1.5) {$O^T$};
    \draw (-2,2) -- (-2,2.75);
    \node[anchor=north] at (-2,2.75) {$|0\rangle_b$};
    \draw (0,0.25) -- (0,1);
    \node[anchor=south] at (0,0.25) {$M_a$};
    \draw (1,0.25) -- (1,1);
    \node[anchor=south] at (1,0.25) {$\alpha$};
    \draw[fill=gray!20] (-0.5,1) rectangle (1.5,2);
    \node at (0.5,1.5) {HKLL$^\dagger$};
\end{scope}

\end{tikzpicture}}}
    = V^\dagger V_\text{HKLL}
\end{equation}
Indeed, equation (\ref{eq:exp_psi1}) applies here, demonstrating that the expectation value of $\tilde\swap_\alpha$ on $V_\alpha |\psi_1\rangle$ agrees with both the bulk ($\swap_\text{baby}$ on $|\psi_1\rangle$) and boundary ($\tilde\swap_\text{AdS}$ on $V|\psi_1\rangle$) SWAP tests. We find that the CO rules also allow the $\alpha$ observer to describe the bulk \textit{with} a baby universe.

Therefore, $\tilde\swap_\alpha$ constructed with the CO rules cannot distinguish between $V_\alpha|\psi_1\rangle$ and $\hat{V}_\alpha|\psi_2\rangle$ when the AR condition (\ref{eq:ARcond}) is satisfied. Whether we use the CO or HUZ rules, the $\alpha$ observer can equally well describe the bulk with or without a baby universe.

\subsection{\texorpdfstring{$\beta$}{beta} observer} \label{sec:CO_beta}
We now turn back to the $\beta$ observer, whose prediction of the SWAP test is given by the reconstruction $\tilde\swap_\beta = R^*_\beta(\swap_\text{baby})$. Recall that the non-isometry of $V_\beta$ may not permit a perfect reconstruction; we will quantify the error in $\langle\tilde\swap_\beta\rangle$ by $\Delta$ defined in (\ref{eq:error}). Since $\langle\swap_\text{baby}\rangle$ is independent of the choice of observer rules, we will reuse (\ref{eq:expSbaby}) when computing $\Delta$.

Let us generalize our ansatz (\ref{eq:S_beta_O}) for the reconstruction $\tilde\swap_\beta$ to include operations on the new boundary degrees of freedom $\partial$ introduced by the CO rules:
\begin{equation}
    \tilde{\swap}_\beta^{(\mathbb{O})} = \tilde{\swap}_\text{AdS} \otimes \mathbb{O}_{\beta'\partial}.
\end{equation}
Leaving $\mathbb{O}$ generic, we calculate its expectation value in $V_\beta|\psi_1\rangle$ for average choices $O'$ in $V_\beta$,
\begin{align*}
    \int dO' \, \langle \tilde{\swap}_\beta^{(\mathbb{O})} \rangle_{V_\beta |\psi_1\rangle} &= \frac{d_{b'}}{d_{b'} + 2} \Bigg[ \tr \bigg( \tr_{\beta\partial} \Big( \big(\psi_\beta \otimes \Phi^+_{\partial\partial'} \big) \mathbb{O}_{\beta\partial} \Big)^2 \bigg) \\
    &\hspace{2.5cm} + \tr \Big( \big(\psi_\beta\otimes \frac{1}{d_\partial}\id_\partial \big) \mathbb{O}_{\beta\partial} \Big)^2 \tr \big( \psi_{1,M_a}^2 \big) \\
    &\hspace{2.5cm} + \tr \bigg( \tr_{\beta\partial} \Big( \big(\psi_\beta\otimes \Phi^+_{\partial\partial'} \big) \mathbb{O}_{\beta\partial} \Big) \tr_{\beta\partial} \Big( \big(\psi_\beta\otimes \Phi^+_{\partial\partial'} \big) \mathbb{O}_{\beta\partial} \Big)^{T} \bigg) \\
    &\hspace{5cm}\times \tr \big( \psi_1 \psi_1^{T_{M_b}} \big) \Bigg],\numberthis \label{eq:SO_CO}
\end{align*}
where $\Phi^+_{\partial\partial'} = |\Phi^+\rangle\langle\Phi^+|_{\partial\partial'}$ is the maximally entangled density matrix $\partial$ and $\partial'$ such that $\tr_{\partial'} \Phi^+_{\partial\partial'} = \id_\partial / d_\partial$. To assist in interpreting this result, figure \ref{fig:SO_line1_CO} provides a circuit diagram of the left hand side, and figure \ref{fig:SO_line2_CO} provides diagrams of the three $\mathbb{O}_{\beta'\partial}$ dependent terms on the right hand side.

\begin{figure}[t]
    \centering
    \begin{tikzpicture}[thick, scale=0.8]


\draw[fill=gray!20] (0.25,-0.5) rectangle (1.25,0.5);
\node[scale=1.2] at (0.75,0) {$\psi_\alpha$};
\draw[fill=gray!20] (1.75,-0.5) rectangle (3.75,0.5);
\node[scale=1.2] at (2.75,0) {$\psi_1$};
\draw[fill=gray!20] (6.25,-0.5) rectangle (7.25,0.5);
\node[scale=1.2] at (6.75,0) {$\psi_\beta$};

\begin{scope}[xscale=-1]
\draw[fill=gray!20] (0.25,-0.5) rectangle (1.25,0.5);
\node[scale=1.2] at (0.75,0) {$\psi_\alpha$};
\draw[fill=gray!20] (1.75,-0.5) rectangle (3.75,0.5);
\node[scale=1.2] at (2.75,0) {$\psi_1$};
\draw[fill=gray!20] (6.25,-0.5) rectangle (7.25,0.5);
\node[scale=1.2] at (6.75,0) {$\psi_\beta$};
\end{scope}

\begin{scope}[yscale=-1]


\draw (0.75,0.5) -- (0.75,1.25);
\draw (2.25,0.5) -- (2.25,1.25);
\draw (3.25,0.5) -- (3.25,1.25);
\draw (6.75,0.5) -- (6.75,4.75);
\draw (4.75,1.25) -- (4.75,1) -- (5.25,0.75) -- (5.75,1) -- (5.75,4.75);
\draw[fill=gray!20] (0.5,1.25) rectangle (2.5,2.25);
\node[scale=1] at (1.5,1.75) {HKLL$^\dagger$};
\draw (1.5,2.25) -- (1.5,3);
\draw[fill=gray!20] (3,1.25) rectangle (5,2.25);
\node[scale=1] at (4,1.75) {$O'^T$};
\draw (4,2.25) -- (4,2.75);
\node[scale=1,anchor=north] at (4,2.75) {$|0\rangle$};
\draw[fill=gray!20] (0.5,3) rectangle (2.5,4);
\node[scale=1] at (1.5,3.5) {HKLL};
\draw (0.75,4) -- (0.75,4.75) -- (0.75,6.25);
\node[scale=0.8,anchor=east] at (0.75,5.875) {$\alpha$};
\draw (2.25,4) -- (2.25,4.75) -- (-2.25,5.5) -- (-2.25,6.25);
\node[scale=0.8,anchor=west] at (2.25,5.875) {$M_a$};
\draw[fill=gray!20] (0.5,6.25) rectangle (2.5,7.25);
\node[scale=1] at (1.5,6.75) {HKLL$^\dagger$};
\draw (1.5,7.25) -- (1.5,8);
\node[scale=1,anchor=north] at (1.5,8) {$A$};
\draw[fill=gray!20] (5.25,4.75) rectangle (7.25,5.75);
\node[scale=1.4] at (6.25,5.25) {$\mathbb{O}$};
\draw (6.75,5.75) -- (6.75,8);
\node[scale=1,anchor=north] at (6.75,8) {$\beta'$};
\draw (5.75,5.75) -- (5.75,8);
\node[scale=1,anchor=north] at (5.75,8) {$\partial$};


\begin{scope}[xscale=-1]
\draw (0.75,0.5) -- (0.75,1.25);
\draw (2.25,0.5) -- (2.25,1.25);
\draw (3.25,0.5) -- (3.25,1.25);
\draw (6.75,0.5) -- (6.75,4.75);
\draw (4.75,1.25) -- (4.75,1) -- (5.25,0.75) -- (5.75,1) -- (5.75,4.75);
\draw[fill=gray!20] (0.5,1.25) rectangle (2.5,2.25);
\node[scale=1] at (1.5,1.75) {HKLL$^\dagger$};
\draw (1.5,2.25) -- (1.5,3);
\draw[fill=gray!20] (3,1.25) rectangle (5,2.25);
\node[scale=1] at (4,1.75) {$O'^T$};
\draw (4,2.25) -- (4,2.75);
\node[scale=1,anchor=north] at (4,2.75) {$|0\rangle$};
\draw[fill=gray!20] (0.5,3) rectangle (2.5,4);
\node[scale=1] at (1.5,3.5) {HKLL};
\draw (0.75,4) -- (0.75,4.75) -- (0.75,6.25);
\node[scale=0.8,anchor=east] at (0.75,5.875) {$\alpha$};
\draw (2.25,4) -- (2.25,4.75) -- (-2.25,5.5) -- (-2.25,6.25);
\node[scale=0.8,anchor=west] at (2.25,5.875) {$M_a$};
\draw[fill=gray!20] (0.5,6.25) rectangle (2.5,7.25);
\node[scale=1] at (1.5,6.75) {HKLL$^\dagger$};
\draw (1.5,7.25) -- (1.5,8);
\node[scale=1,anchor=north] at (1.5,8) {$A$};
\draw[fill=gray!20] (5.25,4.75) rectangle (7.25,5.75);
\node[scale=1.4] at (6.25,5.25) {$\mathbb{O}$};
\draw (6.75,5.75) -- (6.75,8);
\node[scale=1,anchor=north] at (6.75,8) {$\beta'$};
\draw (5.75,5.75) -- (5.75,8);
\node[scale=1,anchor=north] at (5.75,8) {$\partial$};
\end{scope}

\end{scope}

\begin{scope}[yscale=1]


\draw (0.75,0.5) -- (0.75,1.25);
\draw (2.25,0.5) -- (2.25,1.25);
\node[scale=0.8,anchor=east] at (2.25,0.875) {$M_a$};
\draw (3.25,0.5) -- (3.25,1.25);
\node[scale=0.8,anchor=west] at (3.25,0.875) {$M_b$};
\draw (4.75,1.25) -- (4.75,1) -- (5.25,0.75) -- (5.75,1) -- (5.75,3);
\node[scale=0.8,anchor=north] at (4.75,1) {$\partial'$}; 
\draw (6.75,0.5) -- (6.75,3);
\draw[fill=gray!20] (0.5,1.25) rectangle (2.5,2.25);
\node[scale=1] at (1.5,1.75) {HKLL};
\draw (1.5,2.25) -- (1.5,3);
\draw[fill=gray!20] (3,1.25) rectangle (5,2.25);
\node[scale=1] at (4,1.75) {$O'$};
\draw (4,2.25) -- (4,2.75);
\node[scale=1,anchor=south] at (4,2.75) {$\langle0|$};


\begin{scope}[xscale=-1]
\draw (0.75,0.5) -- (0.75,1.25);
\draw (2.25,0.5) -- (2.25,1.25);
\node[scale=0.8,anchor=west] at (2.25,0.875) {$M_a$};
\draw (3.25,0.5) -- (3.25,1.25);
\node[scale=0.8,anchor=east] at (3.25,0.875) {$M_b$};
\draw (4.75,1.25) -- (4.75,1) -- (5.25,0.75) -- (5.75,1) -- (5.75,3);
\node[scale=0.8,anchor=north] at (4.75,1) {$\partial'$}; 
\draw (6.75,0.5) -- (6.75,3);
\draw[fill=gray!20] (0.5,1.25) rectangle (2.5,2.25);
\node[scale=1] at (1.5,1.75) {HKLL};
\draw (1.5,2.25) -- (1.5,3);
\draw[fill=gray!20] (3,1.25) rectangle (5,2.25);
\node[scale=1] at (4,1.75) {$O'$};
\draw (4,2.25) -- (4,2.75);
\node[scale=1,anchor=south] at (4,2.75) {$\langle0|$};
\end{scope}

\end{scope}


\draw[decorate,decoration={brace,amplitude=6pt}] (-8,-0.5) -- (-8,0.5);
\node[scale=1,anchor=east] at (-8.5,0) {$(\psi_1\otimes\psi_\alpha\otimes\psi_\beta)^{\otimes2}$};

\draw[decorate,decoration={brace,amplitude=6pt}] (-8,0.75) -- (-8,2.75);
\node[scale=1,anchor=east] at (-8.5,1.75) {$V_\beta$};

\draw[decorate,decoration={brace,amplitude=6pt}] (-8,-2.75) -- (-8,-0.75);
\node[scale=1,anchor=east] at (-8.5,-1.75) {$V_\beta^\dagger$};

\draw[decorate,decoration={brace,amplitude=6pt}] (-8,-8) -- (-8,-3);
\node[scale=1,anchor=east] at (-8.5,-5.5) {$\tilde\swap_\text{AdS}\otimes\mathbb{O}_{\beta'\partial}$};

\end{tikzpicture}
    \caption{A circuit diagram depiction for $\langle\tilde\swap_\beta^{(\mathbb{O})}\rangle_{V_\beta|\psi_1\rangle}$ appearing in the left hand side of (\ref{eq:SO_CO}). Note that all HKLL operators will drop out after the trace is taken due to the isometry of HKLL.}
    \label{fig:SO_line1_CO}
\end{figure}
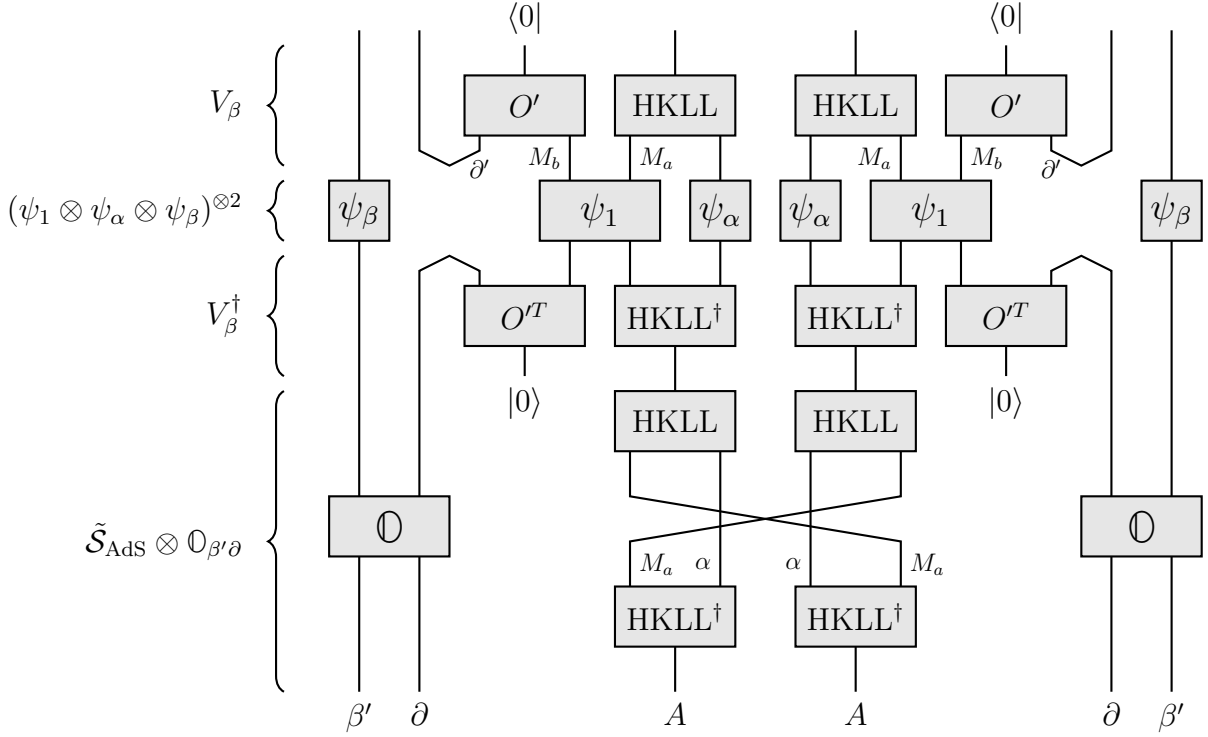

\begin{figure}
    \centering
    \begin{tikzpicture}[thick, scale=1, every node/.style={transform shape}]

\begin{scope}[shift={(0,0)}]
    \draw[fill=gray!20] (2.25,0.5) rectangle (3.25,-0.5);
    \node[scale=1.2] at (2.75,0) {$\psi_\beta$};
    \draw[fill=gray!20] (-2.25,0.5) rectangle (-3.25,-0.5);
    \node[scale=1.2] at (-2.75,0) {$\psi_\beta$};
    \draw (0.75,1.5) -- (0.75,0.5) -- (1.25,0.25) -- (1.75,0.5) -- (1.75,1.5);
    \node[anchor=south west] at (0.75,0.5) {$\partial'$};
    \node[anchor=south west] at (1.75,0.5) {$\partial$};
    \draw (0.75,-1.5) -- (0.75,-0.5) -- (1.25,-0.25) -- (1.75,-0.5) -- (1.75,-1.5) -- (3.5,-1.5) -- (3.5,3.5) -- (1.75,3.5) -- (1.75,2.5);
    \node at (1.25,0) {$\Phi^+$};
    \draw (-0.75,1.5) -- (-0.75,0.5) -- (-1.25,0.25) -- (-1.75,0.5) -- (-1.75,1.5);
    \node[anchor=south east] at (-0.75,0.5) {$\partial'$};
    \node[anchor=south east] at (-1.75,0.5) {$\partial$};
    \draw (-0.75,-1.5) -- (-0.75,-0.5) -- (-1.25,-0.25) -- (-1.75,-0.5) -- (-1.75,-1.5) -- (-3.5,-1.5) -- (-3.5,3.5) -- (-1.75,3.5) -- (-1.75,2.5);
    \node at (-1.25,0) {$\Phi^+$};
    \draw (2.75,0.5) -- (2.75,1.5);
    \draw (2.75,2.5) -- (2.75,3) -- (3.375,3) -- (3.375,-1) -- (2.75,-1) -- (2.75,-0.5);
    \draw (-2.75,0.5) -- (-2.75,1.5);
    \draw (-2.75,2.5) -- (-2.75,3) -- (-3.375,3) -- (-3.375,-1) -- (-2.75,-1) -- (-2.75,-0.5);
    \draw[fill=gray!20] (1.25,1.5) rectangle (3.25,2.5);
    \node[scale=1.4] at (2.25,2) {$\mathbb{O}$};
    \draw[fill=gray!20] (-1.25,1.5) rectangle (-3.25,2.5);
    \node[scale=1.4] at (-2.25,2) {$\mathbb{O}$};
    \draw (0.75,1.5) -- (-0.75,-1.5);
    \draw (-0.75,1.5) -- (0.75,-1.5);
    \node[scale=1,anchor=east] at (-3.75,0) {$\tr \bigg( \tr_{\beta\partial} \Big( \big(\psi_\beta \otimes \Phi^+_{\partial\partial'} \big) \mathbb{O}_{\beta\partial} \Big)^2 \bigg) = $};
\end{scope}

\begin{scope}[shift={(0,-6)}]
    \draw[fill=gray!20] (2.25,0.5) rectangle (3.25,-0.5);
    \node[scale=1.2] at (2.75,0) {$\psi_\beta$};
    \draw[fill=gray!20] (-2.25,0.5) rectangle (-3.25,-0.5);
    \node[scale=1.2] at (-2.75,0) {$\psi_\beta$};
    \draw (0.75,1.5) -- (0.75,0.5) -- (1.25,0.25) -- (1.75,0.5) -- (1.75,1.5);
    \node[anchor=south west] at (0.75,0.5) {$\partial'$};
    \node[anchor=south west] at (1.75,0.5) {$\partial$};
    \draw (0.75,-1.5) -- (0.75,-0.5) -- (1.25,-0.25) -- (1.75,-0.5) -- (1.75,-1.5) -- (3.5,-1.5) -- (3.5,3.5) -- (1.75,3.5) -- (1.75,2.5);
    \node at (1.25,0) {$\Phi^+$};
    \draw (-0.75,1.5) -- (-0.75,0.5) -- (-1.25,0.25) -- (-1.75,0.5) -- (-1.75,1.5);
    \node[anchor=south east] at (-0.75,0.5) {$\partial'$};
    \node[anchor=south east] at (-1.75,0.5) {$\partial$};
    \draw (-0.75,-1.5) -- (-0.75,-0.5) -- (-1.25,-0.25) -- (-1.75,-0.5) -- (-1.75,-1.5) -- (-3.5,-1.5) -- (-3.5,3.5) -- (-1.75,3.5) -- (-1.75,2.5);
    \node at (-1.25,0) {$\Phi^+$};
    \draw (2.75,0.5) -- (2.75,1.5);
    \draw (2.75,2.5) -- (2.75,3) -- (3.375,3) -- (3.375,-1) -- (2.75,-1) -- (2.75,-0.5);
    \draw (-2.75,0.5) -- (-2.75,1.5);
    \draw (-2.75,2.5) -- (-2.75,3) -- (-3.375,3) -- (-3.375,-1) -- (-2.75,-1) -- (-2.75,-0.5);
    \draw[fill=gray!20] (1.25,1.5) rectangle (3.25,2.5);
    \node[scale=1.4] at (2.25,2) {$\mathbb{O}$};
    \draw[fill=gray!20] (-1.25,1.5) rectangle (-3.25,2.5);
    \node[scale=1.4] at (-2.25,2) {$\mathbb{O}$};
    \draw (0.75,1.5) -- (0.125,1.5) -- (0.125,-1.5) -- (0.75,-1.5);
    \draw (-0.75,1.5) -- (-0.125,1.5) -- (-0.125,-1.5) -- (-0.75,-1.5);
    \node[scale=1,anchor=east] at (-3.75,0) {$\tr \Big( \big(\psi_\beta\otimes \frac{1}{d_\partial}\id_\partial \big) \mathbb{O}_{\beta\partial} \Big)^2 = $};
\end{scope}

\begin{scope}[shift={(0,-12)}]
    \draw[fill=gray!20] (2.25,0.5) rectangle (3.25,-0.5);
    \node[scale=1.2] at (2.75,0) {$\psi_\beta$};
    \draw[fill=gray!20] (-2.25,0.5) rectangle (-3.25,-0.5);
    \node[scale=1.2] at (-2.75,0) {$\psi_\beta$};
    \draw (0.75,1.5) -- (0.75,0.5) -- (1.25,0.25) -- (1.75,0.5) -- (1.75,1.5);
    \node[anchor=south west] at (0.75,0.5) {$\partial'$};
    \node[anchor=south west] at (1.75,0.5) {$\partial$};
    \draw (0.75,-1.5) -- (0.75,-0.5) -- (1.25,-0.25) -- (1.75,-0.5) -- (1.75,-1.5) -- (3.5,-1.5) -- (3.5,3.5) -- (1.75,3.5) -- (1.75,2.5);
    \node at (1.25,0) {$\Phi^+$};
    \draw (-0.75,1.5) -- (-0.75,0.5) -- (-1.25,0.25) -- (-1.75,0.5) -- (-1.75,1.5);
    \node[anchor=south east] at (-0.75,0.5) {$\partial'$};
    \node[anchor=south east] at (-1.75,0.5) {$\partial$};
    \draw (-0.75,-1.5) -- (-0.75,-0.5) -- (-1.25,-0.25) -- (-1.75,-0.5) -- (-1.75,-1.5) -- (-3.5,-1.5) -- (-3.5,3.5) -- (-1.75,3.5) -- (-1.75,2.5);
    \node at (-1.25,0) {$\Phi^+$};
    \draw (2.75,0.5) -- (2.75,1.5);
    \draw (2.75,2.5) -- (2.75,3) -- (3.375,3) -- (3.375,-1) -- (2.75,-1) -- (2.75,-0.5);
    \draw (-2.75,0.5) -- (-2.75,1.5);
    \draw (-2.75,2.5) -- (-2.75,3) -- (-3.375,3) -- (-3.375,-1) -- (-2.75,-1) -- (-2.75,-0.5);
    \draw[fill=gray!20] (1.25,1.5) rectangle (3.25,2.5);
    \node[scale=1.4] at (2.25,2) {$\mathbb{O}$};
    \draw[fill=gray!20] (-1.25,1.5) rectangle (-3.25,2.5);
    \node[scale=1.4] at (-2.25,2) {$\mathbb{O}$};
    \draw (0.75,1.5) -- (-0.75,1.5);
    \draw (-0.75,-1.5) -- (0.75,-1.5);
    \node[scale=0.9,anchor=east] at (-3.75,0) {$\tr \bigg( \tr_{\beta\partial} \Big( \big(\psi_\beta\otimes \Phi^+_{\partial\partial'} \big) \mathbb{O}_{\beta\partial} \Big) \tr_{\beta\partial} \Big( \big(\psi_\beta\otimes \Phi^+_{\partial\partial'} \big) \mathbb{O}_{\beta\partial} \Big)^{T} \bigg) = $};
\end{scope}
    
\end{tikzpicture}
    \caption{Circuit diagram depictions of the $\mathbb{O}_{\beta'\partial}$ dependent terms in the right hand side of (\ref{eq:SO_CO}) after the average over $O'$ has been taken.}
    \label{fig:SO_line2_CO}
\end{figure}

Let us now choose $\mathbb{O}_{\beta'\partial}$ to minimize the error $\Delta$ in the expectation value of $\tilde\swap_\beta^{(\mathbb{O})}$. Notice again that the $\psi_1$ dependence is the same in (\ref{eq:expSbaby}) and (\ref{eq:SO_CO}); we therefore minimize $\Delta$ by setting the leading term in (\ref{eq:SO_CO}) to 1,
\begin{equation}
    \tr \bigg( \tr_{\beta\partial} \Big( \big(\psi_\beta \otimes \Phi^+_{\partial\partial'} \big) \mathbb{O}_{\beta\partial} \Big)^2 \bigg) = 1.
\end{equation}
This can be done by choosing $\mathbb{O}_{\beta'\partial} = \tilde\swap_{\beta'\partial}$ to swap the observer $\beta'$ and boundary $\partial$ subsystems.\footnote{Because $\psi_\beta$ is pure, it is actually sufficient to just swap $\partial$ such that $\mathbb{O}_{\beta'\partial} = \id_{\beta'} \otimes \tilde\swap_\partial$. In this case, the $\beta$ observer can perform the SWAP test without having to act on themselves. We have chosen the more general $\mathbb{O}_{\beta'\partial}$ here to parallel (\ref{eq:S_beta_O}) found using the HUZ rules.} 
The expectation value of $\tilde\swap_\beta^{(\swap)}$ defined by this choice is
\begin{equation}
    \int dO' \langle \tilde{\swap}_\beta^{(\swap)} \rangle_{V_\beta |\psi_1\rangle} \approx 1 + \frac{1}{d_\partial} \tr \big( \psi_{1,M_a}^2 \big),
\end{equation}
where we have taken the large $d_b$ limit and dropped the third term as subleading. The error in this reconstruction is given by
\begin{equation} \label{eq:deltaS_CO}
    \overline{\Delta}_\swap \approx \left( 1 - \frac{1}{d_\partial} \right) \tr \big(\psi_{1,M_a}^2\big).
\end{equation}
Had we instead chosen $\mathbb{O}_{\beta'\partial} = \id_{\beta'\partial}$, we would have found an expectation value of 
\begin{equation}
    \int dO' \langle \tilde{\swap}_\beta^{(\id)} \rangle_{V_\beta |\psi_1\rangle} \approx \frac{1}{d_\partial} + \tr \big( \psi_{1,M_a}^2 \big)
\end{equation}
where again we have taken the large $d_b$ limit and dropped the third term as subleading. The error in this choice of reconstruction is given by
\begin{equation} \label{eq:delta1_CO}
    \overline{\Delta}_\id \approx 1 - \frac{1}{d_\partial}.
\end{equation}
Just as we found in (\ref{eq:deltaS}) using the HUZ rules, the error in the expectation value of $\tilde\swap_\beta^{(\swap)}$ is improved compared to $\tilde\swap_\beta^{(\id)}$ by the entanglement between $M_a$ and $M_b$ in $|\psi_1\rangle$.

Finally, note that these average errors are controlled by the dimension $d_\partial$ of the dangling legs between the observer and the rest of the tensor network. Following \cite{akers_observers_2025}, we can generalize $d_\partial$ to be given by the area $A_\partial$ of the boundary between the observer and its environment,
\begin{equation}
    d_\partial \to e^{A_\partial/4G}.
\end{equation}
Therefore, the average error in $\beta$'s description using the CO rules seems to depend on their surface area. Following the lessons learned in section \ref{sec:HUZ_beta} from the HUZ rules, we do not take these errors to be fundamental limitations on $\beta$'s description.

\section{Conclusion} \label{sec:conc}

We have used the observer-modified holographic maps $\hat{V}_\alpha$, $V_\alpha$, and $V_\beta$ introduced by EGH to define new operators for observers $\alpha$ in the AdS bulk and $\beta$ in the baby universe to make predictions for the bulk SWAP test defined in (\ref{eq:S_AdS}). These new operators -- summarized in figure \ref{fig:Op_map} -- allow both observers to describe the baby universe to greater accuracy. For example, the $\alpha$ observer can measure the SWAP test using $\tilde\swap_\alpha$ defined (\ref{eq:S_alpha}). When the AR condition (\ref{eq:ARcond}) is satisfied, we find that $\tilde\swap_\alpha$ cannot distinguish between the bulk states $|\psi_1\rangle$ with a baby universe and $|\psi_2\rangle$ without. This indicates that both are valid bulk states according to $\alpha$. These results -- summarized in table \ref{tab:summary} -- are independent of whether the HUZ or CO rules are used to construct $\hat{V}_\alpha$ and $V_\alpha$. 

\def\arraystretch{1.25}
\begin{table}
    \centering
    \begin{tabular}{c|c|c}
    Which bulk?                 &   EGH \cite{engelhardt_observer_2025} &   This work       \\\hline
    $\alpha$ observer in AdS    &   $|\psi_2\rangle$                    &   $|\psi_1\rangle$ or $|\psi_2\rangle$ \\
    $\beta$ observer in baby    &   $|\psi_1\rangle$, limited accuracy  &   $|\psi_1\rangle$, improved accuracy
\end{tabular}
    \caption{Comparison of the bulk states found to be valid according the AdS observer $\alpha$ and the baby universe observer $\beta$. $|\psi_1\rangle$ represents the bulk state with a baby universe; $|\psi_2\rangle$ represents the bulk state without a baby universe.}
    \label{tab:summary}
\end{table}

For the $\beta$ observer, the non-isometry of $V_\beta$ still generically limits their predictions of the SWAP test on $|\psi_1\rangle$. However, we find that $\tilde\swap_\text{AdS}\otimes\tilde\swap_{\beta'}$ improves on EGH's choice of $\tilde\swap_\text{AdS}\otimes\id_{\beta'}$ for $\beta$'s measurement of the SWAP test. Whether the HUZ or CO rules are used to construct $V_\beta$, the accuracy of the new operator $\tilde\swap_\text{AdS}\otimes\tilde\swap_{\beta'}$ is increased by entanglement between the baby universe and the AdS bulks, consistent with results from \cite{antonini_cosmology_2023,antonini_baby_2025}. These results are summarized in table \ref{tab:beta}. We emphasize that even this improved error is not fundamental; exact reconstructions are possible, and we found a toy example in section \ref{sec:HUZ_beta} where $\beta$'s description was exact.

\begin{table}
    \centering
    \begin{tabular}{c|c|c}
    $\overline\Delta$    &   $\tilde\swap_\text{AdS} \otimes \id_{\beta'}$   &   $\tilde\swap_\text{AdS} \otimes \tilde\swap_{\beta'}$ \\ \hline
    HUZ rules   &   $1 - e^{-S_2(\omega_\beta)}$                    &   $e^{-S_2(\psi_{1,M_a})}( 1 - e^{-S_2(\omega_\beta)})$ \\
    CO rules    &   $1 - \frac{1}{d_\partial}$                      &   $e^{-S_2(\psi_{1,M_a})}( 1 - \frac{1}{d_\partial} )$
    \end{tabular}
    \caption{The average error $\overline\Delta$ in $\beta$'s description defined in (\ref{eq:error}) for two candidate reconstructions of $\swap_\text{baby}$ found using the HUZ rules and CO rules. In either case, we find that the error in the new operator $\tilde\swap_\text{AdS} \otimes \tilde\swap_{\beta'}$ is improved over EGH's choice of $\tilde\swap_\text{AdS} \otimes \id_{\beta'}$ by entanglement between AdS matter $M_a$ and baby universe matter $M_b$ in $|\psi_1\rangle$.}
    \label{tab:beta}
\end{table}

\subsection{Open questions}

Throughout, we have assumed that the AR condition (\ref{eq:ARcond}) is either exactly satisfied or broken by the bulk states $|\psi_1\rangle$ and $|\psi_2\rangle$. More likely, realistic constructions of these states will only satisfy this condition approximately, and we might wonder if this would affect the predictions of either observer. Note that this does not impact equations (\ref{eq:exp_psi2}) and (\ref{eq:exp_psi1}) -- $\tilde\swap_\alpha$ still agrees with $\swap_\text{AdS}$ on $|\psi_2\rangle$ and $\swap_\text{baby}$ on $|\psi_1\rangle$. Furthermore, the error (\ref{eq:error}) between $\tilde\swap_\beta$ and $\swap_\text{baby}$ on $|\psi_1\rangle$ is unchanged, so the accuracy of $\beta$'s description is unaffected. However, a measurement of $\tilde\swap_\alpha$ may now differ slightly between the two states:
\begin{equation}
    \langle\tilde\swap_\alpha\rangle_{\hat{V}_\alpha|\psi_2\rangle} \approx \langle\tilde\swap_\alpha\rangle_{V_\alpha|\psi_1\rangle}.
\end{equation}
How the magnitude of this difference compares with the complexity of $\tilde\swap_\alpha$ should determine whether or not the $\alpha$ observer can use the SWAP test to distinguish between the two bulk states.

Finally, we note that observer complementarity is interesting for more than just the closed universes that we have focused on in this work. For example, EGH also considers the case of an evaporating black hole where interior and exterior observers do not always agree on the entropy of exterior Hawking radiation. Here there are only three holographic maps: one without any observer, one for the exterior observer, and one for the interior observer. It would be interesting to understand if the analysis used in this work could also be used to improve the interior observer's description of the exterior radiation. We leave further investigation of this to future work.

\section*{Acknowledgments}

I am grateful to Chris Akers, Luca Ciambelli, Oliver DeWolfe, Netta Engelhardt, Elliott Gesteau, Daniel Harlow, Chris Waddell, and Beni Yoshida for helpful discussions, and to Alexandre Belin for discussions on related work. Research at Perimeter Institute is supported in part by the Government of Canada through the Department of Innovation, Science and Economic Development and by the Province of Ontario through the Ministry of Colleges and Universities.

No generative AI was used in the conduct of research or at any stage in the preparation of this manuscript.

\bibliographystyle{JHEP}
\bibliography{baby_comp}

@article{engelhardt_observer_2025,
  title = {Observer Complementarity for Black Holes and Holography},
  author = {Engelhardt, Netta and Gesteau, Elliott and Harlow, Daniel},
  year = {2025},
  month = jul,
  number = {arXiv:2507.06046},
  eprint = {2507.06046},
  primaryclass = {hep-th},
  publisher = {arXiv},
  doi = {10.48550/arXiv.2507.06046},
  archiveprefix = {arXiv}
}

@article{higginbotham_tests_2025,
  title = {On Tests for Baby Universes in {{AdS}}/{{CFT}}},
  author = {Higginbotham, Kenneth},
  year = {2025},
  month = sep,
  journal = {Journal of High Energy Physics},
  volume = {2025},
  number = {9},
  pages = {38},
  issn = {1029-8479},
  doi = {10.1007/JHEP09(2025)038}
}

@article{antonini_cosmology_2023,
  title = {Cosmology from Random Entanglement},
  author = {Antonini, Stefano and Sasieta, Martin and Swingle, Brian},
  year = {2023},
  month = nov,
  number = {arXiv:2307.14416},
  eprint = {2307.14416},
  primaryclass = {hep-th},
  publisher = {arXiv},
  doi = {10.48550/arXiv.2307.14416},
  archiveprefix = {arXiv}
}

@article{harlow_quantum_2025,
  title = {Quantum Mechanics and Observers for Gravity in a Closed Universe},
  author = {Harlow, Daniel and Usatyuk, Mykhaylo and Zhao, Ying},
  year = {2025},
  month = jan,
  number = {arXiv:2501.02359},
  eprint = {2501.02359},
  primaryclass = {hep-th},
  publisher = {arXiv},
  doi = {10.48550/arXiv.2501.02359},
  archiveprefix = {arXiv}
}

@article{akers_observers_2025,
  title = {On Observers in Holographic Maps},
  author = {Akers, Chris and Bueller, Gracemarie and DeWolfe, Oliver and Higginbotham, Kenneth and Reinking, Johannes and Rodriguez, Rudolph},
  year = {2025},
  month = may,
  journal = {Journal of High Energy Physics},
  volume = {2025},
  number = {5},
  pages = {201},
  issn = {1029-8479},
  doi = {10.1007/JHEP05(2025)201}
}

@article{antonini_holographic_2025,
  title = {Do Holographic {{CFT}} States Have Unique Semiclassical Bulk Duals?},
  author = {Antonini, Stefano and Rath, Pratik},
  year = {2025},
  month = mar,
  number = {arXiv:2408.02720},
  eprint = {2408.02720},
  primaryclass = {hep-th},
  publisher = {arXiv},
  doi = {10.48550/arXiv.2408.02720},
  archiveprefix = {arXiv}
}

@article{engelhardt_further_2025,
  title = {Further {{Evidence Against}} a {{Semiclassical Baby Universe}} in {{AdS}}/{{CFT}}},
  author = {Engelhardt, Netta and Gesteau, Elliott},
  year = {2025},
  month = apr,
  number = {arXiv:2504.14586},
  eprint = {2504.14586},
  primaryclass = {hep-th},
  publisher = {arXiv},
  doi = {10.48550/arXiv.2504.14586},
  archiveprefix = {arXiv}
}

@article{akers_reconstruction_2025,
  title = {On the Reconstruction Map in {{JT}} Gravity},
  author = {Akers, Chris and Lucas, Andrew and Vikram, Amit},
  year = {2025},
  month = jun,
  number = {arXiv:2506.18975},
  eprint = {2506.18975},
  primaryclass = {hep-th},
  publisher = {arXiv},
  doi = {10.48550/arXiv.2506.18975},
  archiveprefix = {arXiv}
}

@article{harlow_gauging_2023,
  title = {Gauging Spacetime Inversions in Quantum Gravity},
  author = {Harlow, Daniel and Numasawa, Tokiro},
  year = {2023},
  month = nov,
  number = {arXiv:2311.09978},
  eprint = {2311.09978},
  primaryclass = {hep-th},
  publisher = {arXiv},
  doi = {10.48550/arXiv.2311.09978},
  archiveprefix = {arXiv}
}

@article{hamilton_holographic_2006,
  title = {Holographic Representation of Local Bulk Operators},
  author = {Hamilton, Alex and Kabat, Daniel and Lifschytz, Gilad and Lowe, David A.},
  year = 2006,
  month = sep,
  journal = {Physical Review D},
  volume = {74},
  number = {6},
  eprint = {hep-th/0606141},
  pages = {066009},
  issn = {1550-7998, 1550-2368},
  doi = {10.1103/PhysRevD.74.066009},
  archiveprefix = {arXiv}
}

@article{hamilton_local_2006,
  title = {Local Bulk Operators in {{AdS}}/{{CFT}}: A Boundary View of Horizons and Locality},
  shorttitle = {Local Bulk Operators in {{AdS}}/{{CFT}}},
  author = {Hamilton, Alex and Kabat, Daniel and Lifschytz, Gilad and Lowe, David A.},
  year = 2006,
  month = apr,
  journal = {Physical Review D},
  volume = {73},
  number = {8},
  eprint = {hep-th/0506118},
  pages = {086003},
  issn = {1550-7998, 1550-2368},
  doi = {10.1103/PhysRevD.73.086003},
  archiveprefix = {arXiv}
}

@article{hamilton_local_2007,
  title = {Local Bulk Operators in {{AdS}}/{{CFT}}: {{A}} Holographic Description of the Black Hole Interior},
  shorttitle = {Local Bulk Operators in {{AdS}}/{{CFT}}},
  author = {Hamilton, Alex and Kabat, Daniel and Lifschytz, Gilad and Lowe, David A.},
  year = 2007,
  month = may,
  journal = {Physical Review D},
  volume = {75},
  number = {10},
  eprint = {hep-th/0612053},
  pages = {106001},
  issn = {1550-7998, 1550-2368},
  doi = {10.1103/PhysRevD.75.106001 10.1103/PhysRevD.75.129902},
  archiveprefix = {arXiv}
}

@article{gesteau_nogo_2025,
  title = {A No-Go Theorem for Large \${{N}}\$ Closed Universes},
  author = {Gesteau, Elliott},
  year = 2025,
  month = sep,
  number = {arXiv:2509.14338},
  eprint = {2509.14338},
  primaryclass = {hep-th},
  publisher = {arXiv},
  doi = {10.48550/arXiv.2509.14338},
  archiveprefix = {arXiv}
}

@article{kudler-flam_emergent_2025,
  title = {Emergent {{Mixed States}} for {{Baby Universes}} and {{Black Holes}}},
  author = {{Kudler-Flam}, Jonah and Witten, Edward},
  year = 2025,
  month = oct,
  number = {arXiv:2510.06376},
  eprint = {2510.06376},
  primaryclass = {hep-th},
  publisher = {arXiv},
  doi = {10.48550/arXiv.2510.06376},
  archiveprefix = {arXiv}
}

@article{liu_holographic_2025,
  title = {Towards a Holographic Description of Closed Universes},
  author = {Liu, Hong},
  year = 2025,
  month = sep,
  number = {arXiv:2509.14327},
  eprint = {2509.14327},
  primaryclass = {hep-th},
  publisher = {arXiv},
  doi = {10.48550/arXiv.2509.14327},
  archiveprefix = {arXiv}
}

@article{maldacena_Wormholes_2004,
  title = {Wormholes in {{AdS}}},
  author = {Maldacena, Juan and Maoz, Liat},
  year = {2004},
  month = mar,
  journal = {Journal of High Energy Physics},
  volume = {2004},
  number = {02},
  pages = {053},
  issn = {1126-6708},
  doi = {10.1088/1126-6708/2004/02/053},
  langid = {english}
}

@article{almheiri_Page_2020,
  title = {The {{Page}} Curve of {{Hawking}} Radiation from Semiclassical Geometry},
  author = {Almheiri, Ahmed and Mahajan, Raghu and Maldacena, Juan and Zhao, Ying},
  year = {2020},
  month = mar,
  journal = {Journal of High Energy Physics},
  volume = {2020},
  number = {3},
  eprint = {1908.10996},
  primaryclass = {hep-th},
  pages = {149},
  issn = {1029-8479},
  doi = {10.1007/JHEP03(2020)149},
  archiveprefix = {arXiv},
  langid = {english}
}

@article{penington_Replica_2020,
  title = {Replica Wormholes and the Black Hole Interior},
  author = {Penington, Geoff and Shenker, Stephen H. and Stanford, Douglas and Yang, Zhenbin},
  year = {2020},
  month = apr,
  number = {arXiv:1911.11977},
  eprint = {1911.11977},
  primaryclass = {hep-th},
  publisher = {arXiv},
  doi = {10.48550/arXiv.1911.11977},
  archiveprefix = {arXiv},
  langid = {english}
}

@article{marolf_Transcending_2020,
  title = {Transcending the Ensemble: Baby Universes, Spacetime Wormholes, and the Order and Disorder of Black Hole Information},
  shorttitle = {Transcending the Ensemble},
  author = {Marolf, Donald and Maxfield, Henry},
  year = {2020},
  month = aug,
  journal = {Journal of High Energy Physics},
  volume = {2020},
  number = {8},
  pages = {44},
  issn = {1029-8479},
  doi = {10.1007/JHEP08(2020)044},
  langid = {english}
}

@article{mcnamara_Baby_2020,
  title = {Baby {{Universes}}, {{Holography}}, and the {{Swampland}}},
  author = {McNamara, Jacob and Vafa, Cumrun},
  year = {2020},
  month = aug,
  number = {arXiv:2004.06738},
  eprint = {2004.06738},
  primaryclass = {hep-th},
  publisher = {arXiv},
  doi = {10.48550/arXiv.2004.06738},
  archiveprefix = {arXiv}
}

@article{usatyuk_Closed_2024,
  title = {Closed Universes in Two Dimensional Gravity},
  author = {Usatyuk, Mykhaylo and Wang, Zi-Yue and Zhao, Ying},
  year = {2024},
  month = aug,
  journal = {SciPost Physics},
  volume = {17},
  number = {2},
  pages = {051},
  issn = {2542-4653},
  doi = {10.21468/SciPostPhys.17.2.051},
  langid = {english}
}

@article{usatyuk_Closed_2025,
  title = {Closed Universes, Factorization, and Ensemble Averaging},
  author = {Usatyuk, Mykhaylo and Zhao, Ying},
  year = {2025},
  month = feb,
  journal = {Journal of High Energy Physics},
  volume = {2025},
  number = {2},
  pages = {52},
  issn = {1029-8479},
  doi = {10.1007/JHEP02(2025)052},
  langid = {english}
}

@article{abdalla_Gravitational_2025,
  title = {The Gravitational Path Integral from an Observer's Point of View},
  author = {Abdalla, Ahmed I. and Antonini, Stefano and Iliesiu, Luca V. and Levine, Adam},
  year = {2025},
  month = jan,
  number = {arXiv:2501.02632},
  eprint = {2501.02632},
  primaryclass = {hep-th},
  publisher = {arXiv},
  doi = {10.48550/arXiv.2501.02632},
  archiveprefix = {arXiv}
}

@article{susskind_stretched_1993,
  title = {The {{Stretched Horizon}} and {{Black Hole Complementarity}}},
  author = {Susskind, L. and Thorlacius, L. and Uglum, J.},
  year = 1993,
  month = oct,
  journal = {Physical Review D},
  volume = {48},
  number = {8},
  eprint = {hep-th/9306069},
  pages = {3743--3761},
  issn = {0556-2821},
  doi = {10.1103/PhysRevD.48.3743},
  archiveprefix = {arXiv}
}

@article{maldacena_Large_1999,
  title = {The {{Large N Limit}} of {{Superconformal Field Theories}} and {{Supergravity}}},
  author = {Maldacena, Juan M.},
  year = {1999},
  journal = {International Journal of Theoretical Physics},
  volume = {38},
  number = {4},
  eprint = {hep-th/9711200},
  pages = {1113--1133},
  issn = {00207748},
  doi = {10.1023/A:1026654312961},
  archiveprefix = {arXiv}
}

@article{gubser_Gauge_1998,
  title = {Gauge {{Theory Correlators}} from {{Non-Critical String Theory}}},
  author = {Gubser, S. S. and Klebanov, I. R. and Polyakov, A. M.},
  year = {1998},
  month = may,
  journal = {Physics Letters B},
  volume = {428},
  number = {1-2},
  eprint = {hep-th/9802109},
  pages = {105--114},
  issn = {03702693},
  doi = {10.1016/S0370-2693(98)00377-3},
  archiveprefix = {arXiv}
}

@article{witten_Sitter_1998,
  title = {Anti de {{Sitter}} Space and Holography},
  author = {Witten, Edward},
  year = {1998},
  month = jan,
  journal = {Advances in Theoretical and Mathematical Physics},
  volume = {2},
  number = {2},
  pages = {253--291},
  publisher = {International Press of Boston},
  issn = {1095-0753},
  doi = {10.4310/ATMP.1998.v2.n2.a2},
  langid = {english}
}

@article{chen_Observers_2025,
  title = {Observers Seeing Gravitational {{Hilbert}} Spaces: Abstract Sources for an Abstract Path Integral},
  shorttitle = {Observers Seeing Gravitational {{Hilbert}} Spaces},
  author = {Chen, Hong Zhe},
  year = {2025},
  month = may,
  number = {arXiv:2505.15892},
  eprint = {2505.15892},
  primaryclass = {hep-th},
  publisher = {arXiv},
  doi = {10.48550/arXiv.2505.15892},
  archiveprefix = {arXiv},
  langid = {english}
}

@article{hayden_holographic_2016,
	title = {Holographic duality from random tensor networks},
	volume = {2016},
	issn = {1029-8479},
	url = {http://arxiv.org/abs/1601.01694},
	doi = {10.1007/JHEP11(2016)009},
	number = {11},
	urldate = {2024-01-24},
	journal = {Journal of High Energy Physics},
	author = {Hayden, Patrick and Nezami, Sepehr and Qi, Xiao-Liang and Thomas, Nathaniel and Walter, Michael and Yang, Zhao},
	month = nov,
	year = {2016},
	note = {arXiv:1601.01694 [cond-mat, physics:hep-th, physics:math-ph, physics:quant-ph]},
	pages = {9}
}

@article{antonini_baby_2025,
  title = {The {{Baby Universe}} Is {{Fine}} and the {{CFT Knows It}}: {{On Holography}} for {{Closed Universes}}},
  shorttitle = {The {{Baby Universe}} Is {{Fine}} and the {{CFT Knows It}}},
  author = {Antonini, Stefano and Rath, Pratik and Sasieta, Martin and Swingle, Brian and L{\'o}pez, Alejandro Vilar},
  year = 2025,
  month = aug,
  number = {arXiv:2507.10649},
  eprint = {2507.10649},
  primaryclass = {hep-th},
  publisher = {arXiv},
  doi = {10.48550/arXiv.2507.10649},
  archiveprefix = {arXiv}
}

@article{verstraete_renormalization_2004,
  title = {Renormalization Algorithms for {{Quantum-Many Body Systems}} in Two and Higher Dimensions},
  author = {Verstraete, F. and Cirac, J. I.},
  year = 2004,
  month = jul,
  number = {arXiv:cond-mat/0407066},
  eprint = {cond-mat/0407066},
  publisher = {arXiv},
  doi = {10.48550/arXiv.cond-mat/0407066},
  archiveprefix = {arXiv}
}

@article{akers_background_2024,
  title = {Background Independent Tensor Networks},
  author = {Akers, Chris and Wei, Annie Y.},
  year = 2024,
  month = sep,
  journal = {SciPost Physics},
  volume = {17},
  number = {3},
  eprint = {2402.05910},
  primaryclass = {hep-th},
  pages = {090},
  issn = {2542-4653},
  doi = {10.21468/SciPostPhys.17.3.090},
  archiveprefix = {arXiv}
}

@article{mori_baby_2025,
  title = {Baby Universe as Logical Qubits: Information Recovery in Random Encoding},
  shorttitle = {Baby Universe as Logical Qubits},
  author = {Mori, Takato and Yoshida, Beni},
  year = 2025,
  month = nov,
  number = {arXiv:2511.20747},
  eprint = {2511.20747},
  primaryclass = {hep-th},
  publisher = {arXiv},
  doi = {10.48550/arXiv.2511.20747},
  archiveprefix = {arXiv}
}

@article{dewolfe_nonisometric_2023,
  title = {Non-Isometric Codes for the Black Hole Interior from Fundamental and Effective Dynamics},
  author = {DeWolfe, Oliver and Higginbotham, Kenneth},
  year = 2023,
  month = sep,
  journal = {Journal of High Energy Physics},
  volume = {2023},
  number = {9},
  pages = {68},
  issn = {1029-8479},
  doi = {10.1007/JHEP09(2023)068}
}

@article{dewolfe_bulk_2024,
  title = {Bulk Reconstruction and Non-Isometry in the Backwards-Forwards Holographic Black Hole Map},
  author = {DeWolfe, Oliver and Higginbotham, Kenneth},
  year = 2024,
  month = jun,
  journal = {Journal of High Energy Physics},
  volume = {2024},
  number = {6},
  pages = {126},
  issn = {1029-8479},
  doi = {10.1007/JHEP06(2024)126}
}

@article{sasieta_baby_2025,
  title = {Baby {{Universes}} from {{Thermal Pure States}} in {{SYK}}},
  author = {Sasieta, Martin and Swingle, Brian and L{\'o}pez, Alejandro Vilar},
  year = 2025,
  month = nov,
  number = {arXiv:2512.00149},
  eprint = {2512.00149},
  primaryclass = {hep-th},
  publisher = {arXiv},
  doi = {10.48550/arXiv.2512.00149},
  archiveprefix = {arXiv}
}

@article{liu_filtering_2025,
  title = {"{{Filtering}}" {{CFTs}} at Large {{N}}: {{Euclidean Wormholes}}, {{Closed Universes}}, and {{Black Hole Interiors}}},
  shorttitle = {"{{Filtering}}" {{CFTs}} at Large {{N}}},
  author = {Liu, Hong},
  year = 2025,
  month = dec,
  number = {arXiv:2512.13807},
  eprint = {2512.13807},
  primaryclass = {hep-th},
  publisher = {arXiv},
  doi = {10.48550/arXiv.2512.13807},
  archiveprefix = {arXiv}
}

@article{belin_baby_2025,
  title = {Baby {{Universes}} in {{AdS}}\$\_3\$},
  author = {Belin, Alexandre and de Boer, Jan},
  year = 2025,
  month = dec,
  number = {arXiv:2512.02098},
  eprint = {2512.02098},
  primaryclass = {hep-th},
  publisher = {arXiv},
  doi = {10.48550/arXiv.2512.02098},
  archiveprefix = {arXiv}
}

\end{document}